%% file: main.tex
\documentclass[aps,prl,showpacs,preprintnumbers,amsmath,amssymb,superscriptaddress,twocolumn,longbibliography]{revtex4-1} %
\usepackage{amsmath}
\usepackage{graphicx}
\usepackage{float}
\usepackage{ulem}
\usepackage{physics}
\usepackage{comment}
\usepackage{xcolor}
\definecolor{gold}{rgb}{0.83, 0.69, 0.22}

\newcommand\COMMENTED[1] {}

\newcommand{\ssec}[1]{\textit{#1}.---}
\newcommand{\bs}{\boldsymbol}
\newcommand{\up}{\uparrow}
\newcommand{\dn}{\downarrow}

\newcommand{\cmat}{c_{ak}}
\newcommand{\ansatz}{(MP)$^2$NQS}

\newcommand{\nematic}{NSCL} %

\newcommand{\nematics}{\nematic~}

\graphicspath{{./figures/}{./figures_supp/}{./NNDensities/}}

\begin{document}
\title{Ground state phases of the two-dimension electron gas with a unified variational approach}
\author{Conor Smith}
\thanks{These authors contributed equally.}
\affiliation{Center for Computational Quantum Physics, Flatiron Institute, New York, NY, 10010, USA}
\affiliation{Department of Electrical and Computer Engineering, University of New Mexico, Albuquerque, NM 87131, USA}
\author{Yixiao Chen}
\thanks{These authors contributed equally.}
\affiliation{Center for Computational Quantum Physics, Flatiron Institute, New York, NY, 10010, USA}
\affiliation{Program in Applied and Computational Mathematics, Princeton University, Princeton, New Jersey, 08544, USA}

\author{Ryan Levy}
\affiliation{Center for Computational Quantum Physics, Flatiron Institute, New York, NY, 10010, USA}
\author{Yubo Yang}
\affiliation{Center for Computational Quantum Physics, Flatiron Institute, New York, NY, 10010, USA}
\author{Miguel A. Morales}
\affiliation{Center for Computational Quantum Physics, Flatiron Institute, New York, NY, 10010, USA}
\author{Shiwei Zhang}
\affiliation{Center for Computational Quantum Physics, Flatiron Institute, New York, NY, 10010, USA}
\date{\today}
\begin{abstract}
The two-dimensional electron gas (2DEG) is a fundamental model,
which is drawing increasing interest because of recent 
advances in experimental and theoretical studies of 2D materials. Current understanding of the 
ground state of the 2DEG relies on quantum Monte Carlo 
calculations, based on variational comparisons of different 
ansatze for different phases. We use a single variational 
ansatz, 
a general backflow-type wave function using 
a message-passing neural quantum    state architecture, for a unified description across the entire density range. The variational optimization consistently
leads to lower %
ground-state energies than previous best results. %
Transition into a Wigner crystal (WC) phase occurs 
automatically %
at $r_s=37\pm 1$,
a density lower than currently believed. %
Between the liquid and WC phases, 
the same ansatz and variational search strongly suggest the existence of intermediate states in a broad range of densities, with
enhanced short-range nematic spin correlations.
\end{abstract}
\maketitle

The %
uniform electron gas,
also known as jellium, %
has been one of the 
most important fundamental models in 
condensed matter physics.
The ground-state energies computed in jellium
\cite{ceperley_ground_1980,tanatar_ground_1989,attaccalite_correlation_2002,drummond_phase_2009} also serve as the foundation
for density functionals which are
widely applied in {\it ab initio\/} computations in many disciplines.
Recently the 
two-dimensional electron gas (2DEG) has
drawn renewed interest owing to the many recent experimental discoveries in 
two-dimensional materials~\cite{hossain_observation_2020,hossain_spontaneous_2021,hossain_anisotropic_2022,smolenski_signatures_2021,falson_competing_2022}.
Despite the simplicity of the model, the phase 
diagram of jellium is not fully understood. %
Theoretically, a 
variety of candidate ground states
have been proposed \cite{falakshahi_hybrid_2005}, including theory of microemulsion~\cite{Spivak2004} and metalic electron crystal~\cite{KSKim2023}
between phase transitions.
Computationally, the most accurate 
studies have confirmed only 
a Fermi liquid (FL) phase at high density, and a Wigner crystal (WC) phase
(possibly with different spin polarizations) at low density, 
with a transition between them estimated
at $r_s=31 \pm 1$ in two dimensions, based on the most accurate computations 
to date \cite{drummond_phase_2009}. Experimentally, 
generalized 
WC phases, in the presence of external potentials,
have been detected \cite{li_imaging_2021,shabani_deep_2021,regan_mott_2020}, and 
higher density regimes are being 
realized and studied. In particular,
a low-temperature magneto-optical experiment in MoSe$_2$ \cite{Sung2023}
reported evidence for a liquid phase with enhanced spin susceptibility and a microemulsion phase in between the WC and the FL.

The primary challenge in computationally determining the ground-state phases in jellium is that candidate states are separated by extremely small energy differences. In order to make further progress, we need methods that have 
systematically improvable and, possibly more importantly, balanced accuracy across different regimes. Quantum Monte Carlo calculations have been 
the standard tool used so far.
However, they typically rely on different variational ansatze for different phases.
The ground state at each density is determined 
by comparing the computed 
energies from each candidate 
ansatz, and the search for different states is limited
by the expressiveness of the variational ansatze that can be computed given the available resources.

Recently, neural network wavefunctions~\cite{carleo2017solving,pfau_ab_2020,hermann2020deep,hermann_ab_2023}, also known as neural quantum states (NQS), have been used as ansatze in variational Monte Carlo (VMC) calculations. 
They %
significantly expand the expressiveness of the variational 
wave functions that can be explored and computed. Many developments and applications have 
shown 
the %
promise of
this approach. 
Impressive accuracy 
has been achieved 
in a variety of benchmark studies in fermion systems.
However, 
to date, few 
applications in this area have 
demonstrated the potential predictive power 
that can lead to physical results or insights beyond the reach of existing capabilities.

In this paper, %
we leverage
NQS to study the ground state phases of the 2DEG.
We uncover evidence of short-range nematic spin correlations in a broad range of intermediate densities, 
and automatically discover a ``floating'' WC.
Building on an ansatz of Slater-Jastrow-backflow form with message-passing graph neural network components~\cite{pescia_neural-network_2022},
we introduce multiple planewaves (MPs) into each orbital, allowing the ansatz to span 
both the liquid and WC phases.
Using this single ansatz, 
which we call \ansatz, 
we obtain variational ground-state energies 
significantly lower than those from state-of-the-art QMC~\cite{drummond_phase_2009}. The transition to the WC is 
automatic%
, yielding a lower critical density 
($r_s=37\pm 1$)
than previously 
believed.
In the WC phase, the ansatz yields a ``floating'' crystal
with %
partially restored %
translational symmetry.
With the same ansatz, 
our calculations suggest 
the existence of an intermediate state
in a wide range of densities ($r_s \sim 10$-$35$) before the WC transition, a liquid state with enhanced short-range
spin correlations %
which break rotational symmetry. Although the calculations
are still in modest system sizes (%
but comparable to the largest sizes that 
have been treated by NQS), our analysis indicates that the nematic spin correlation is quite robust, and that the 2DEG may have much new physics to offer.

The 2DEG consists of interacting electrons in a uniform compensating charge background. The 
Hamiltonian (in Hartree atomic units)
\begin{equation}
H = -\dfrac{1}{2}\sum\limits_i\nabla_i^2+
\sum\limits_{i<j}
\dfrac{1}{\vert \bs{r}_i-\bs{r}_j \vert} + {\rm b.g.}
\end{equation}
is specified by a single dimensionless parameter, $r_s$, 
the Wigner-Seitz radius
in units of the Bohr radius $a_\mathrm{B}$
(see considerations of a finite simulation with periodic boundary conditions in SM).
Our \ansatz~ansatz has Slater-Jastrow-backflow form%
\begin{equation}
\Psi(R) =
\text{det}  \left[ 
\{\langle \bs{x}_j; \sigma_j|
 \phi_a; \chi_a\rangle \}\right]
\exp\big( U_2(R)+\mathcal{U}(R) \big)\,,
\label{eq:ansatz-0}
\end{equation}
where $R\equiv \{\bs{r}_1;\sigma_1;
\bs{r}_2;\sigma_2; \cdots; \bs{r}_N;\sigma_N\}$ denotes the  coordinates and spins of the $N$ electrons in the simulation cell of area $\Omega$
($\pi r_s^2=\Omega/N$), 
and the ``quasi-particle'' coordinates are 
\begin{equation}
\label{eq:NN1}
\bs{x}_j \equiv \bs{r}_j +
\mathcal{N}(R)\,.    
\end{equation}
Both $\mathcal{N}$ 
in Eq.~(\ref{eq:NN1})
and $\mathcal{U}$ in the Jastrow in Eq.~(\ref{eq:ansatz-0}) are many-body neural network functions, inspired by the message-passing neural quantum states (MP-NQS) ansatz~\cite{pescia2023message}.
In this work, we restrict the system
to total $S_z=0$, and each
$\sigma_j$ is fixed 
($+1$ or $-1$). %
We also keep 
each orbital $\chi_a$ as an $\hat s_z$ eigenstate. 
(This reduces the Slater determinant 
to a product of separate spin-$\uparrow$
and spin-$\downarrow$ components.)
Each of the $N$ orbitals in the Slater determinant %
can be written as 
\begin{equation}
\label{eq:MPs}
\phi_a(\bs{x}) =  \sum_{k=1}^{N_k} \cmat  \exp(\mathrm{i}\bs{G}_k\cdot\bs{x})\,
\end{equation}
where $\{\cmat\}$ are 
 variational parameters and we choose $N_k \gg N$ (multiple planewaves).
The function $U_2$ 
in Eq.~(\ref{eq:ansatz-0})
is a sum of isotropic two-body correlators:
$U_2(R)=\sum_{j\ne l} u_{\sigma_j\sigma_l}(\bs{x}_j-\bs{x}_l)$
, where $u_{+1}$ and $u_{-1}$ are
standard functions satisfying the cusp 
condition for same and opposite spins, respectively~\cite{ceperley1977monte}.

While the original idea of backflow was based on physical intuition from liquid helium~\cite{feynman_atomic_1954,feynman_energy_1956} and perturbation theory~\cite{kwon_effects_1993,holzmann_backflow_2003,taddei_iterative_2015}, recent implementations~\cite{ruggeri_nonlinear_2018,pfau_ab_2020,cassella_discovering_2023,pescia_neural-network_2022,wilson_neural_2023} parameterize a general all-body, instead of few-body, coordinate transformation.
In this work, we adapt the message passing network from~\cite{pescia2023message} as our many-body backflow transformation %
$\mathcal{N}(R)$,
with modifications to facilitate optimization.
We introduce an extra term %
$\mathcal{U}(R)$ 
in the Jastrow factor that shares the same message-passing layers with the backflow $\mathcal{N}$, to further improve the expressivity of the ansatz.
Details of our implementation can be found in the Supplementary Materials.

We variationally minimize the total energy
\begin{equation}
\label{eq:var-E}
E=\frac{\int  [H\Psi(R)/\Psi(R)]\,|\Psi(R)|^2\,dR}
{\int  |\Psi(R)|^2\,dR}\,.
\end{equation}
The Metropolis-adjusted Langevin algorithm (MALA) ~\cite{besag1994comments} is employed to sample electron configurations 
$\{ R \}$
from the unnormalized distribution %
$|\Psi(R)|^2$,
which allows the evaluation of Eq.~(\ref{eq:var-E}) as the expectation of the 
local energy: 
$\mathbb{E}[H\Psi(R)/\Psi(R)]$.
New configurations in the Markov chain are proposed using overdamped Langevin dynamics, which utilizes gradient information to increase the acceptance rate. 
The step size of the Langevin dynamics is adaptively tuned during the optimization according to the average acceptance rate.
The optimization of the parameters is performed using the stochastic reconfiguration (SR) method~\cite{sorella1998green, sorella2005wave} with recently introduced modifications~\cite{goldshlager2024kaczmarz} to stabilize and accelerate the process.

We apply the same ansatz and optimization procedure to 
different $r_s$
values spanning both liquid and crystal phases. 
The hyperparameters used for different $r_s$ values are the same, except for the initial learning rates, which are tuned for each $r_s$ to account for the different energy scales.
The calculations are primarily carried out with $N=56$ electrons under periodic boundary conditions, 
with spot checks
using $N=58$ (square simulation cell) and $N=120$.

\begin{figure}[ht]
\includegraphics[width=\linewidth]{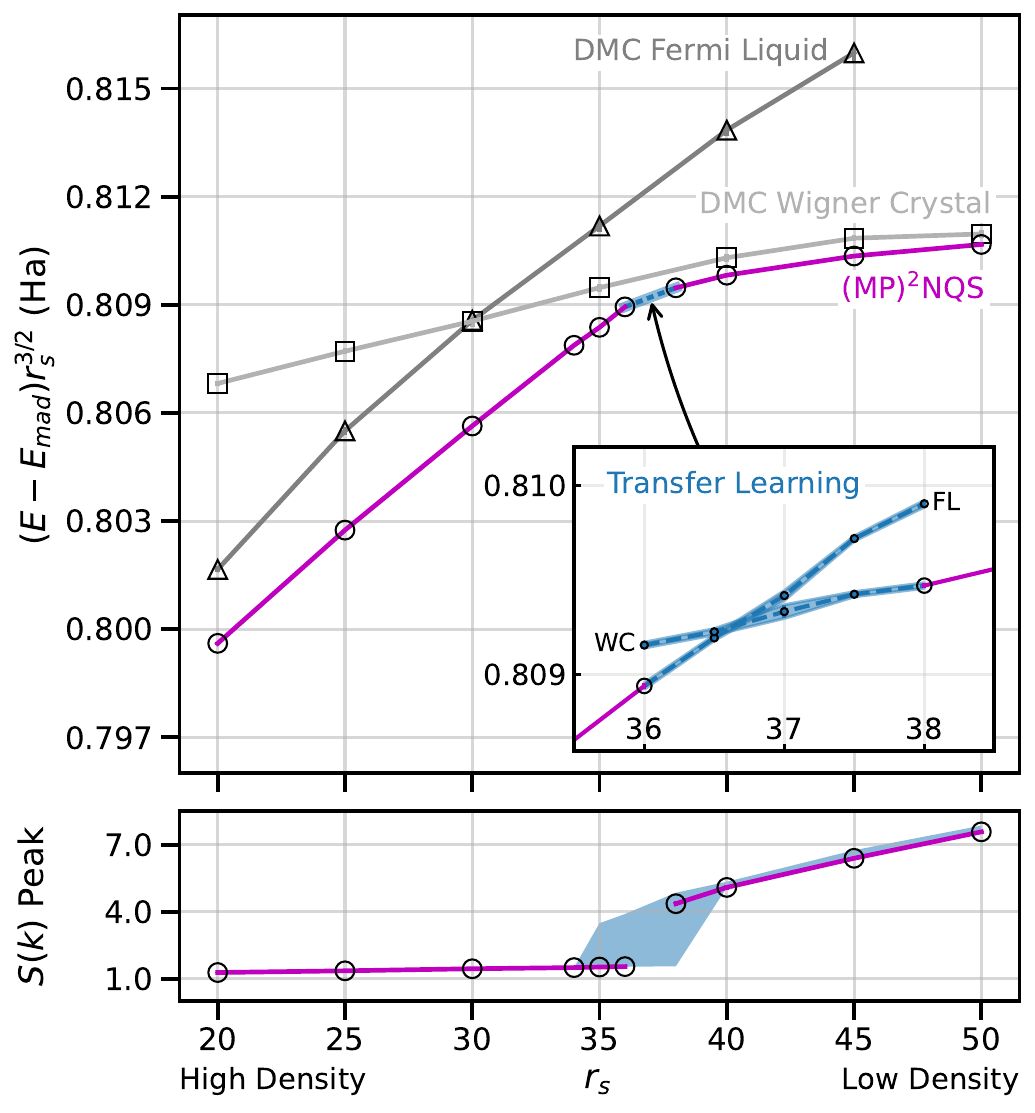}
\caption{
Accurate and automatic detection of 
the Wigner crystal %
transition. 
The ground-state energies 
from variational optimization of the unified \ansatz\ ansatz systematically outperform
state-of-the-art results from 
diffusion Monte Carlo (DMC) using separate variational ansatze
as trial wave functions (top panel).
Statistical error of energies from the \ansatz\ ansatz is smaller than the line width.
The inset
shows a study %
near the WC transition using transfer learning,  
which preserves the character of the 
state 
while changing its density.
The bottom panel shows 
the Bragg peak values in the static structure factor $S(\bs{k})$ calculated from 
\ansatz.
The shaded region indicate uncertainty from the optimization process
as discussed in the main text.
Results here are obtained from simulation cells containing $N=56$ electrons.
}
\label{fig:etot}
\end{figure}

\begin{figure}[ht]
\includegraphics[width=\linewidth]{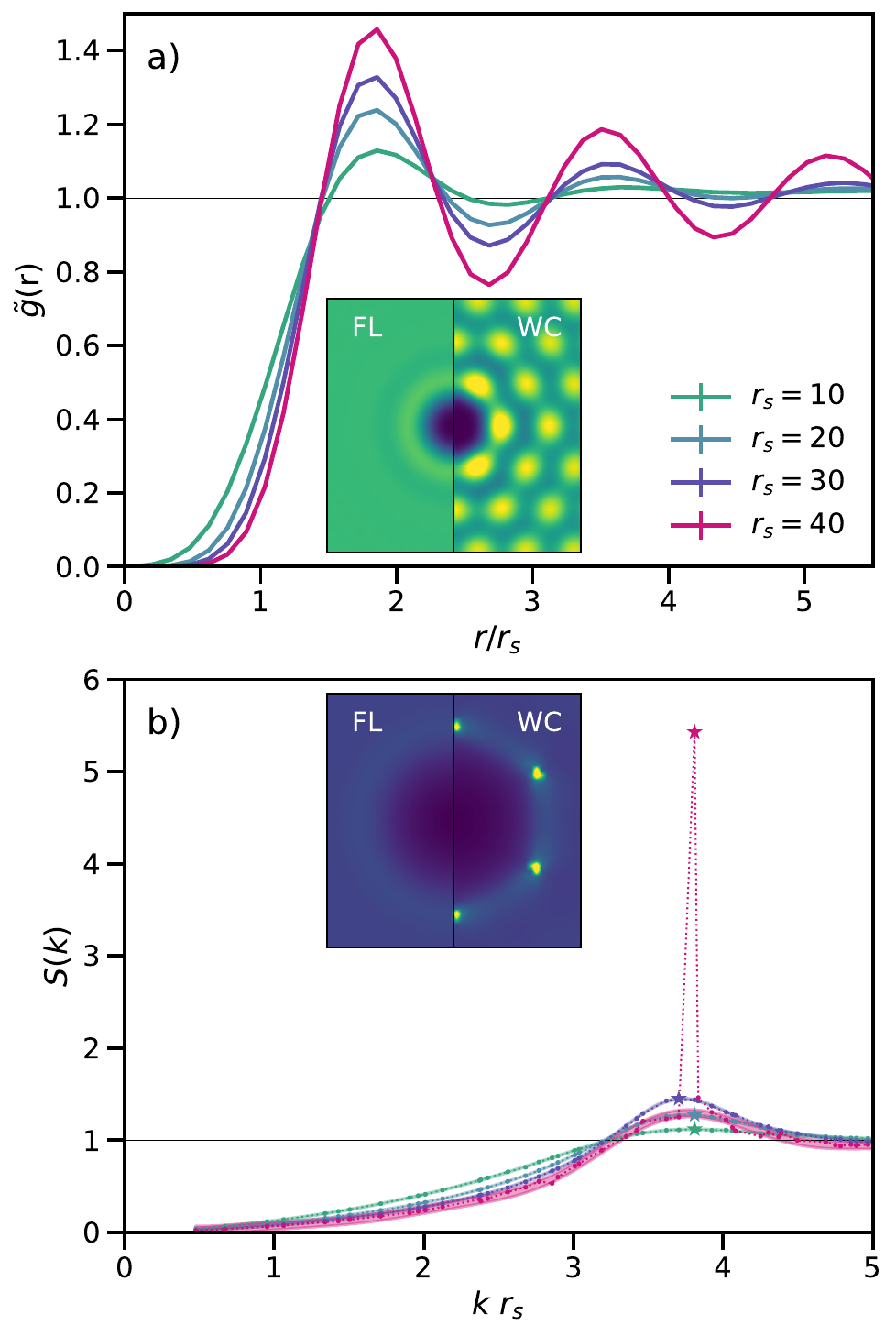}
\caption{
Charge-charge correlations for representative densities, shown as
(a) the %
pair correlation function
$\tilde{g}(\bs{r})$, %
and (b) %
the static structure factor $S(\bs{k})$. %
The main panels show the radially averaged quantities, while the insets show the full two dimensional versions.
Each inset is divided into two halves, with the left showing $r_s=10$ and the right showing $r_s=40$.
At $r_s=40$, long-range correlation is evident from the persistent oscillations in the tail of $\tilde{g}(r)$ and the corresponding Bragg peak (pink star) in $S(k)$.
Results shown in the insets are %
from %
$N=120$. %
}
\label{fig:pair}
\end{figure}

The \ansatz\ ansatz %
automatically and accurately %
discovers the %
WC transition by variational optimization.
Our current 
knowledge of this 
transition is based
on diffusion Monte Carlo (DMC) calculations using different VMC ansatze, 
as shown in Fig.~\ref{fig:etot}. 
The reference DMC calculations~\cite{dmc_details} are performed in identical systems as in the \ansatz\ calculations and obtained with Slater-Jastrow-backflow trial wavefunctions with short-range isotropic two-body correlators, using planewaves for the FL phase and Gaussians for the WC phase. %

The optimized variational energy from our %
neural network ansatz yields a single equation 
of state line shown in the main plot in Fig.~\ref{fig:etot}, which
is lower  than that of DMC at all densities. The difference in energies is especially prominent in the liquid phase.
In the bottom panel of Fig.~\ref{fig:etot}, we can see the crystallinity of the \ansatz\  ansatz spontaneously increase across the WC transition.
Near the transition, in the range $r_s \in (34, 40)$, the ansatz discovers states with and without charge order in a narrow energy window depending on the randomly generated initial parameters. 
The shaded region encapsulates results of all states having energy close to the lowest state (with a threshold of $0.01\%$ of the total energy).
To locate the 
critical $r_s$
more precisely, we use the transfer learning technique to preserve the character 
(FL or WC) of the ansatz in the transition region and calculate its variational energy.
As shown in the inset of Fig.~\ref{fig:etot}, we transferred the liquid state found at $r_s=36$ up to $38$ and the crystal state found at $r_s=38$ down to $r_s=36$.
The energy crossing gives a %
transition point of $r_s=36.5$.
After applying finite-size corrections using energies from
DMC~\cite{drummond_phase_2009} (which is less costly to run for larger simulation sizes),
we obtain an estimated critical value for 
the WC transition:
$r_s=37\pm1$.

The main advantage of the \ansatz\ is that, within the same ansatz, correlated phases such as the WC can be automatically discovered.
Further, once an NQS has been optimized, many physical observables can be calculated to high precision to characterize the discovered phase.
DMC has been the 
standard for 
accurate characterization of 
ground-state properties in jellium.
There have been years of
 sophisticated development and optimization 
 since the early landmark calculation~\cite{ceperley_ground_1980}. 
That the variational energy
is lower than 
DMC is important,
not so much for 
the quantitative 
improvement, but 
as a strong indication of 
the quality of the \ansatz\ wave function.

We next examine 
the properties of the \ansatz\ ground state as a function 
of $r_s$.
We compute 
spin-resolved
one- 
and two-body density functions: %
$\rho_\alpha(\bs{r})=\langle \Psi|\hat \rho_\alpha(\bs{r})|\Psi\rangle$
and 
$\rho^{\alpha,\beta}_2(\bs{r}, \bs{r}')=\langle \Psi|\hat \rho_\alpha(\bs{r})\hat \rho_\beta(\bs{r}')|\Psi\rangle$, 
where $\alpha$ and $\beta$ are spin indices ($\in \Bqty{\uparrow, \downarrow}$)
and
$\hat \rho_\alpha(\bs{r})$ is the density operator at $\bs{r}$ for spin $\alpha$.
The %
uniform pair correlation function is 
\begin{equation} \label{eq:g0ofr}
\tilde{g}(\bs{r}) = %
\frac{\Omega}{N(N-1)}
\int \dd{\bs{r}_0} \sum_{\alpha,\beta \in \{\up,\dn\}}\rho^{\alpha,\beta}_2(\bs{r}_0, \bs{r}_0+\bs{r})\,,
\end{equation}
and the Fourier transform of $\tilde{g}(\bs{r})$ is the structure factor
\begin{align} \label{eq:sofk}
S(\bs{k})
= &~ 1 + %
\frac{N-1}{\Omega}
\int \dd{\bs{r}} e^{-\mathrm{i} \bs{k} \cdot \bs{r} } (\tilde{g}(\bs{r}) - 1)\,.
\end{align}
Figure~\ref{fig:pair}(a) and (b) show %
$\tilde{g}(\bs{r})$ and $S(\bs{k})$
in direct and reciprocal space, respectively.
At $r_s=40$, 
$\tilde{g}(r)$ shows strong and persistent oscillation at large distances, indicative of long-range correlation.
Correspondingly, the $S(\bs{k})$ exhibits Bragg peaks at wave vectors consistent with a triangular WC.
As shown in Fig.~\ref{fig:pair}(b) and the bottom panel of Fig.~\ref{fig:etot}, we %
find 
$S(\bs{k})$ is smooth vs.~$k\equiv |\bs{k}|$ 
for $r_s\le 34$, confirming the lack of long-range charge correlation, 
and the %
character of a liquid.

\begin{figure*}[ht]
\includegraphics[width=\linewidth]{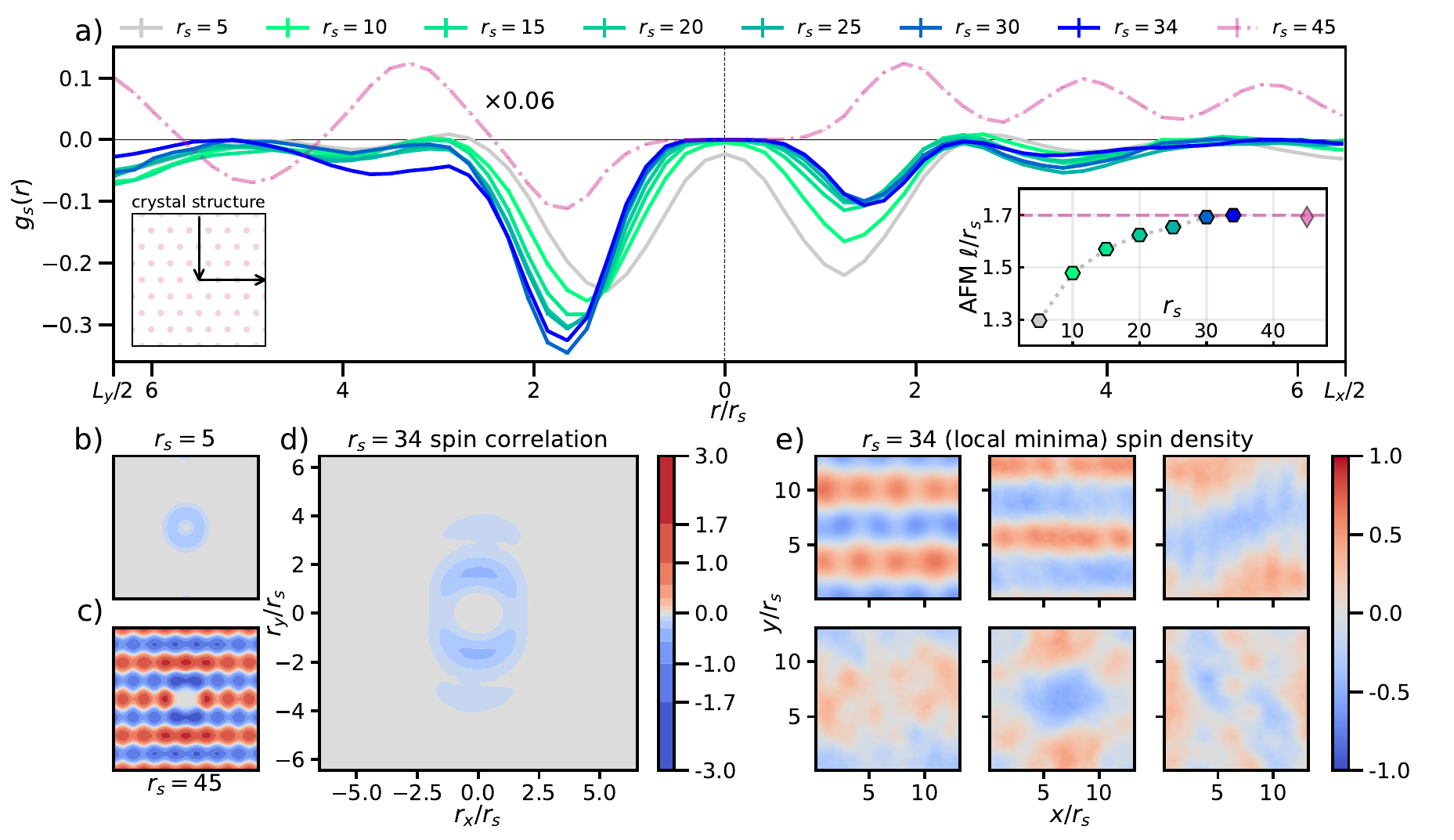}
\caption{Signatures of the intermediate 
\nematics states.
In (a), line cuts of the spin correlation function, $g_s(\bs{r})$, are shown
scanning the densities between a prototypical liquid ($r_s=5$) and a WC %
($r_s=45$).
Significant short-range anisotropic %
spin correlations
are present starting from $r_s \sim 10$. The anisotropy 
grows with $r_s$, with the nearest spin showing stronger (anti-parallel) correlation and drifting further toward the near-neighbor position in the WC, as shown 
in the inset,
which plots the location of the first minimum position in the left half of the main panel, as a function of $r_s$.
The 2D $g_s(\bs{r})$ of representative states are shown in (b) $r_s=5$, 
(c) %
$r_s=45$
and (d) 
$r_s=34$.
At $r_s=34$, we found many local minima having total energy within $0.01$\% of the lowest state.
The spin densities of the six lowest-energy states are shown in (e), %
ordered by their energies starting  
with the lowest at top left, whose 
$g_s(\bs{r})$ is shown in (d), to %
the highest at bottom right of the panel.
}
\label{fig:iftdsk}
\end{figure*}

The WC phase discovered by \ansatz\ exhibits a partially floating nature, characterized by nearly uniform density along the spin stripes of an anti-ferromagnetic (AFM) pattern, as shown in appendix Fig.~\ref{fig:evodens}(d). 
(The  broken symmetry spin stripe pattern is built into the ansatz, because we 
restrict it to colinear spins with 
fixed $\sigma_j$ and
$\chi$ in Eq.~(\ref{eq:ansatz-0}).)
This is different from the usual WC state studied computationally in the 2DEG, in which
a WC is assumed and the electron locations in the crystal are specified {\it a priori\/}. 
This ``floating crystal" state, despite being theoretically predicted long ago~\cite{bishop1982electron,lewin2019floating}, has not been observed in numerical simulations reported in the literature to the best of our knowledge.

One of the long-standing questions on the electron gas is whether there are intermediate states 
between the FL and WC and, if so, of what character.
The \ansatz\ provides a unique opportunity to explore this question, given that the ansatz does not 
require postulating the candidate phases
and that it is capable of high accuracy across the density regimes,
as the results above indicate. 
Below, we investigate the ground state from 
\ansatz\ more closely at the intermediate densities. 
We compute the spin-spin correlation function defined as
\begin{align}
\label{eq:spingofr}
g_s(\bs{r}) = \sum_{\alpha,\beta \in \{\up,\dn\}} ( 2 \delta_{\alpha\beta} - 1 ) g_{\alpha\beta}(\bs{r}),
\end{align}
where $g_{\alpha\beta}(\bs{r})$ is the pair correlation function
$\rho^{\alpha,\beta}_2(\bs{r}_0, \bs{r}_0+\bs{r})/[\rho_\alpha(\bs{r}_0)\rho_\beta(\bs{r}_0+\bs{r})]$ after averaging with respect to the reference position 
$\bs{r}_0$ (as done in Eq.~(\ref{eq:g0ofr})).

We find a strong tendency for short-range anisotropic spin correlations 
in a wide range of intermediate densities, which we will refer to below as 
nematic spin correlated liquid 
(\nematic, for short).
In a typical liquid at $r_s=5$, $g_s(\bs{r})$ exhibits isotropic short-range AFM correlation due to exchange effects, as shown in Fig.~\ref{fig:iftdsk}(b) and the gray line in Fig.~\ref{fig:iftdsk}(a). 
In a typical crystal at $r_s=45$, $g_s(\bs{r})$ has long-range AFM correlation commensurate with the triangular lattice as shown in Fig.~\ref{fig:iftdsk}(c) %
(note stripe 
structure due to restricting to colinear spins, as
discussed earlier)
and the dot-dashed line in Fig.~\ref{fig:iftdsk}(a).
As the interaction strength $r_s$ increases in the liquid phase, the spin channel develops nematic correlation while the charge channel remains isotropic (Fig.~\ref{fig:evogofr}).
The spin correlation remains very short-ranged until the WC is reached; however,
starting from $r_s\sim 10$, a preferential AFM direction emerges.
The AFM correlation strengthens along 
this direction and weakens perpendicular to it.
With $N=56$,  the simulation cell is rectangular, and 
almost all of the \nematics found here 
picked a direction within $10^\circ$ of the %
short axis.
We have also verified with %
calculations %
in the $N=58$ square cell, 
where the AFM is equally likely to be along 
the $x$- or $y$-direction,
and the $g_s(\bs{r})$ is quantitatively consistent with those in Fig.~\ref{fig:iftdsk}(a)
(see appendix Figs.~\ref{fig:supp_afm_rs}
and~\ref{fig:square}).
By following the first dip in $g_s(r)$ along the preferred %
direction, which is aligned to $y$ in Fig.~\ref{fig:iftdsk}(a), we see that the characteristic length of this correlation increases from $\sim 1.5\,r_s$ at $r_s=10$ to $\sim 1.7\,r_s$ around the WC transition, %
as shown in the inset in panel~a.
This is consistent with the projection of the nearest neighbor distance in the WC along the $y$ direction (panel~c).
Therefore, the \nematics connects spin features of the liquid to that of the WC. %

We also observed faint spin density stripes in the \nematic.
As shown in Fig.~\ref{fig:iftdsk}(e), the 
six lowest-energy states found at $r_s=34$ (%
within $0.01\%$ in variational energy)
all show some spin density wave order.
The six states have spin densities, 
$s(\bs{r})\equiv\rho_\uparrow(\bs{r})-\rho_\downarrow(\bs{r})$, which vary 
in pattern,
orientation, and strength.
However they all have $g_s(\bs{r})$
nearly identical to Fig.~\ref{fig:iftdsk}(d), oriented in the direction of the AFM stripes.
The spin density waves in the two lowest-energy states at $r_s=34$ have twice the wavelength of the WC.
Strictly speaking, the broken symmetry in a one-body observable here is artificial, as we discussed earlier in connection with the partially floating crystals. 
We interpret these spin density waves as originating from the anisotropic two-body spin correlation, but pinned to a one-body density, possibly by a rough optimization landscape.
The charge density remains nearly uniform in the \nematics (Fig.~\ref{fig:evodens}).
At $r_s=34$, all the low-energy states we find are liquids.
In contrast, at $r_s=35$, there are some
local minima with WC character very close in energy to the lowest-energy state, which is a liquid (Fig.~\ref{fig:supp_fig4slices}).
Whether this can be a manifestation of microemulsion \cite{Spivak2004} in the small simulation cells is an interesting question to explore.
In principle, the current ansatz allows this and the possible existence of a metallic electronic crystal to be studied.
As shown in Fig.~\ref{fig:iftdsk}(a), the short-range AFM feature is fully developed by $r_s=15$ and persists all the way up to crystallization.
Interestingly, the density range coincides with the
recent experimental observation~\cite{Sung2023} of an intermediate liquid state with high spin susceptibility. %

In summary, we report several advances in characterizing and understanding the 
ground state of the 2DEG, using a neural quantum state ansatz, \ansatz.
The single ansatz 
yields  
variational 
energies which are significantly lower than the current state-of-the-art from 
DMC %
projections of separately optimized VMC wave  functions. 
It allows a unified description of the 2DEG, 
finding the FL and WC phases without any {\it a priori\/} specification,
and revising the transition to the WC to a lower density than  currently assumed. %
The same variational ansatz %
suggests the existence of a correlated liquid state with 
short-range nematic spin correlations at a wide range of intermediate densities.
Further studies are needed, for example 
with more systematic investigation of finite-size effects and more 
extensive 
quantification of the different correlations and spin response, 
in order to better characterize these states and understand their connection with the experimental observations.
But these results  %
from the \ansatz\ %
already provide evidence and interesting clue that there is %
rich physics between the FL and WC phases in the electron gas.

{\it Note added:\/} During the preparation of this manuscript, 
Azadi \textit{et al.}~\cite{Azadi2024} reported 
improved DMC calculations in the liquid phase, and a revised transition density of $r_s=35\pm1$, more
consistent with our result. 

\ssec{Acknowledgment}
The Flatiron Institute is a division of the Simons Foundation. The authors would like to thank 
David Ceperley,
Eugene Demler,
Ilya Esterlis,
Gil Goldshlager, 
Andrew Millis, 
Vadim Oganesyan, 
Agnes Valenti, 
Han Wang, 
Lei Wang, 
Hao Xie, 
Linfeng Zhang
for helpful discussions. 

\bibliography{ref}

\clearpage
\onecolumngrid
\renewcommand{\thefigure}{S\arabic{figure}}
\setcounter{figure}{0}
\include{supp}

\end{document}

%% file: supp.tex
\section{Supplemental Materials: \\
Ground state phases of the two-dimension electron gas \\
with a unified variational approach}

\author{Conor Smith}
\thanks{These authors contributed equally.}
\affiliation{Center for Computational Quantum Physics, Flatiron Institute, New York, NY, 10010, USA}
\affiliation{Department of Electrical and Computer Engineering, University of New Mexico, Albuquerque, NM 87131, USA}
\author{Yixiao Chen}
\thanks{These authors contributed equally.}
\affiliation{Center for Computational Quantum Physics, Flatiron Institute, New York, NY, 10010, USA}
\affiliation{Program in Applied and Computational Mathematics, Princeton University, Princeton, New Jersey, 08544, USA}

\author{Ryan Levy}
\affiliation{Center for Computational Quantum Physics, Flatiron Institute, New York, NY, 10010, USA}
\author{Yubo Yang}
\affiliation{Center for Computational Quantum Physics, Flatiron Institute, New York, NY, 10010, USA}
\author{Miguel A. Morales}
\affiliation{Center for Computational Quantum Physics, Flatiron Institute, New York, NY, 10010, USA}
\author{Shiwei Zhang}
\affiliation{Center for Computational Quantum Physics, Flatiron Institute, New York, NY, 10010, USA}

\ \ 

\ \ 

\ \

\subsection{Total Energy and Static Structure Factor}
 
\begin{figure}[!htbp]
\begin{minipage}{0.48\textwidth}
\includegraphics[width=\linewidth]{{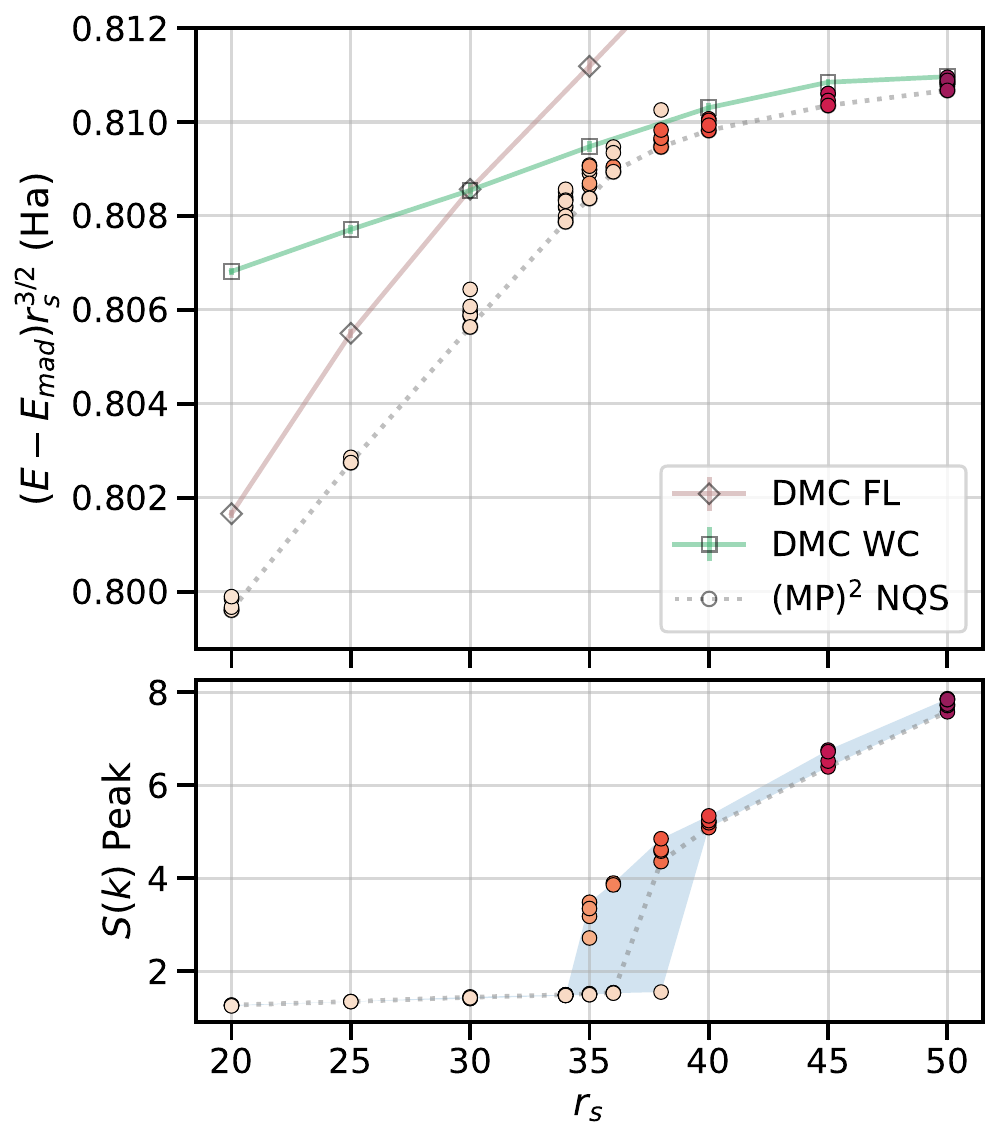}}
(a) Total energy and averaged $S(k)$ peak vs $r_s$.
\end{minipage}
\begin{minipage}{0.48\textwidth}
\includegraphics[width=\linewidth]{{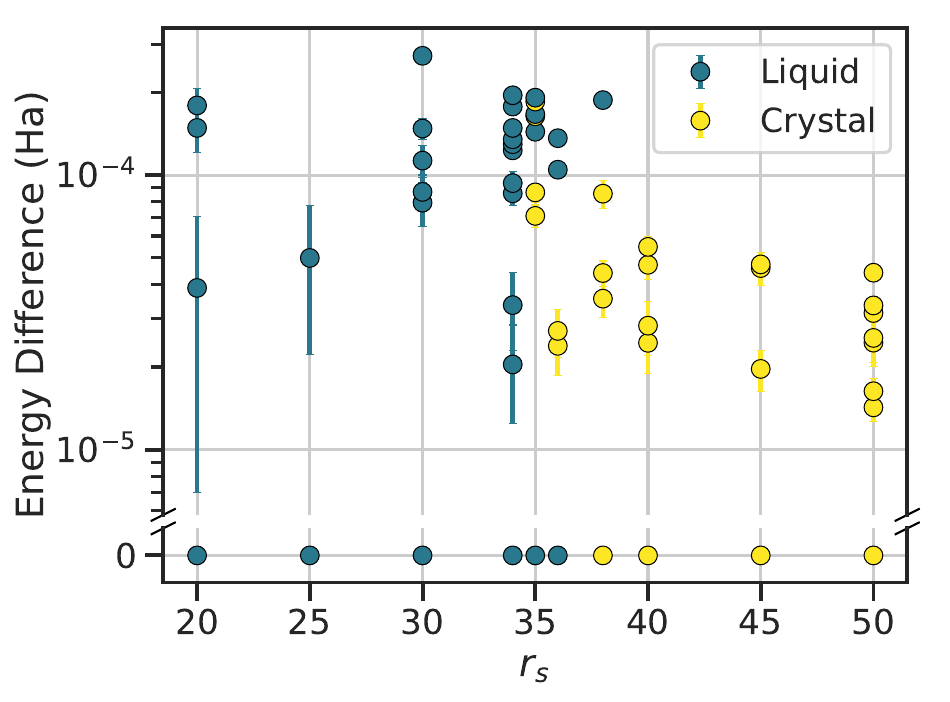}}
(b) Energy difference to the lowest obtained state vs $r_s$.
\end{minipage}
\caption{(a) Compiled optimization trials showing energy (\textit{top}) and $S(k)$ peak (\textit{bottom}) across different values of $r_s$. Grey open circles and a dotted line mark the lowest energy trials for a given density, while a lighter/darker color corresponds to a smaller/larger $S(k)$ peak. Shaded region corresponds to same of fig 1. %
(b) Energy difference to the lowest obtained state vs $r_s$. The different colors denote liquid states vs those that are WC.}
\label{fig:supp_fig4slices}
\end{figure}

In Table~\ref{tab:supp_energy_table}, we provide the lowest total energies and corresponding $S(k)$ Bragg peaks obtained using the \ansatz~ansatz along with total energies of reference DMC calculations.
To emphasize the
variation of final states from the frustrated optimization landscape, we show two figures of merit, the energy and averaged Bragg peaks of the structure factor $S(k)$, in Fig.~\ref{fig:supp_fig4slices}. 
We find many competitive states in the region $r_s\in [34,40)$, depicted both in the energetics and the variety of $S(k)$ Bragg peak values. The lowest energy states in this regime are all liquid like until $r_s=38$, but the optimization may find a crystal or stripe state nearby in energy. Not pictured, there are non-crystalline states found at $r_s \geq 40$ but they are significantly higher than energy, a factor of approximately 3 higher. We note that the Wigner crystal states with slightly higher energies %
still show no signs of defects.
They may have small modulations in the densities and are not as ``floating'' as the lowest energy ones.

\begin{table}[!htbp]
\caption{Total energy per electron $E/N$ in Hartree and average Bragg peak $S_b$ of $56$ electrons at various densities.}
\label{tab:supp_energy_table}
\begin{tabular}{r | l l l | l l}
\hline\hline
$r_s$ & $E/N$ & $S_b$ && $E^\mathrm{DMC}_\mathrm{FL}/N$ & $E^\mathrm{DMC}_\mathrm{WC}/N$ \\
\hline
20 & -0.04637(2) & 1.273(7) && -0.046342(1) & -0.0462847(9) \\
25 & -0.03782(1) & 1.35(1) && -0.0378001(8) &-0.0377824(7) \\
30 & -0.03197(1) & 1.442(6) && -0.0319493(6) & -0.031949(1) \\
34 & -0.028457(6) & 1.493(6) &&  & \\
35 & -0.027699(7) & 1.519(8) && -0.0276854(8) & -0.0276936(5) \\
36 & -0.026980(4) & 1.539(5) &&  & \\
38 & -0.025652(3) & 4.4(1) &&  & \\
40 & -0.024451(2) & 5.1(1)  && -0.0244356(7) & -0.0244496(4)\\
45 & -0.021896(2) & 6.39(8) && -0.0218769(7) & -0.0218940(3)\\
50 & -0.019829(1) & 7.58(6) &&  & -0.0198283(3)\\
\hline\hline
\end{tabular}
\end{table}

\subsection{Evolution of the AFM Correlations }
\begin{figure}[!htbp]
\includegraphics[width=0.5\linewidth]{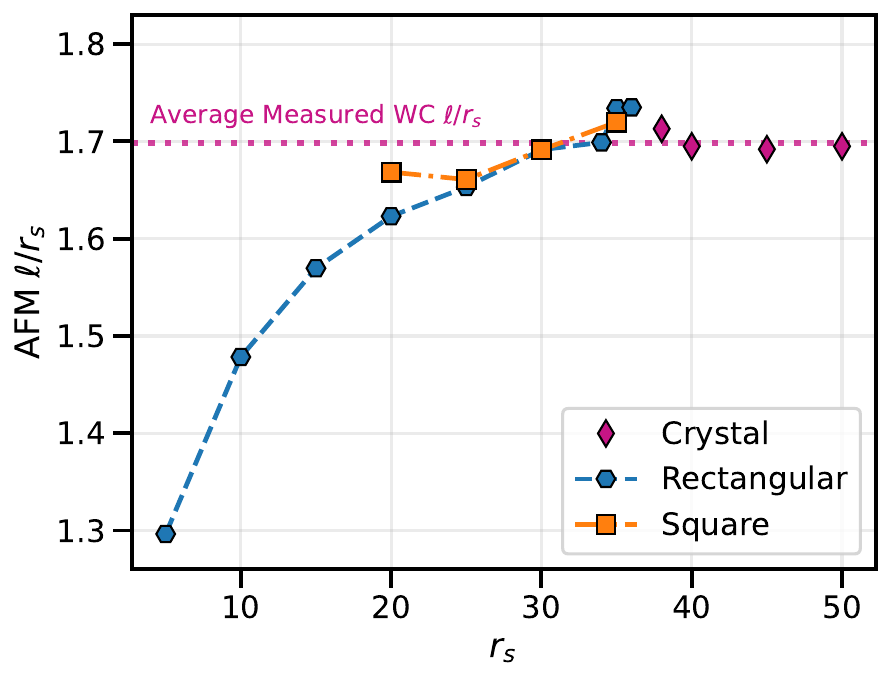}

\caption{Fitted values of the peak location of $g_s(\bs{r})$ along the AFM direction as a function of $r_s$ for both rectangular and square cell of liquid phase and rectangular cell of crystal phase. The dotted horizontal line shows the averaged value of measured WC distance.}
\label{fig:supp_afm_rs}
\end{figure}

In Fig.~\ref{fig:supp_afm_rs}, we explore the peak of the $g_s(\bs{r})$ location as a function of $r_s$ which was originally shown in the linecut of Fig.~\ref{fig:iftdsk}(c). The peak location roughly represents the distance between two pairs of opposite spin electrons, or the AFM characteristic length $\ell$. 
To obtain the peak location, we fit the underlying histogram data to a Gaussian
\begin{equation}
G(r, \theta; \{ \ell, \theta_i, \sigma_r, \sigma_\theta, A, B\}) = A\exp\left( -\dfrac{(r-\ell)^2}{\sigma_r^2} - \dfrac{(\theta-\theta_o)^2}{\sigma_\theta^2} \right) + B
\end{equation}
to mitigate bias from finite grid resolution and statistical noise. As $r_s$ increases, the value of $\ell/r_s$ slowly climbs until $g_s(r)$ abruptly changes when the system becomes a WC at sufficiently large $r_s$.

We also measure the peak location of the triangular WC along the AFM direction (y axis in the simulation cell). The peak location is determined by a spline fit for the data on the grid along that axis. The average measured peak distance for WC is $\ell/r_s \approx 1.70$, as shown in the dotted horizontal line in Fig.~\ref{fig:supp_afm_rs}. This value is slightly larger than what is predicted from a perfect triangular lattice, $\frac{\sqrt{3}}{2} \times \sqrt{\frac{2\pi}{\sqrt{3}}} \approx 1.65$, since the correlation function is distorted locally by the exchange effects.

We also note that the peak distance we get for $r_s = 35$ and $36$ are slightly higher than the averaged crystal value, by around 2 \%. One potential cause for this small overshoot is the frustrated optimization landscape, since we have multiple competing local minima of different phases within a small energy range at these two $r_s$ (see Fig.~\ref{fig:supp_fig4slices}). Whether there is some physical explanation (such as microemulsion) behind this phenomenon is a direction to explore in the future.

\subsection{Evolution of observables}
\begin{figure}[!htbp]
\begin{minipage}{0.48\textwidth}
\includegraphics[width=\linewidth]{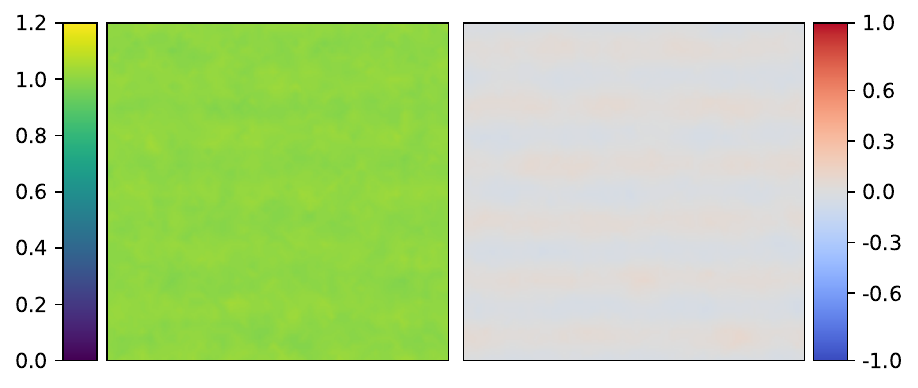}
(a) $r_s = 5$
\end{minipage}
\begin{minipage}{0.48\textwidth}
\includegraphics[width=\linewidth]{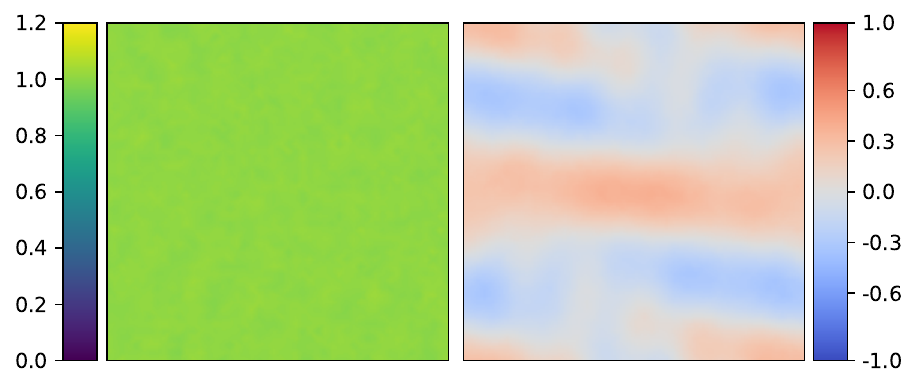}
(b) $r_s=20$
\end{minipage}
\begin{minipage}{0.48\textwidth}
\includegraphics[width=\linewidth]{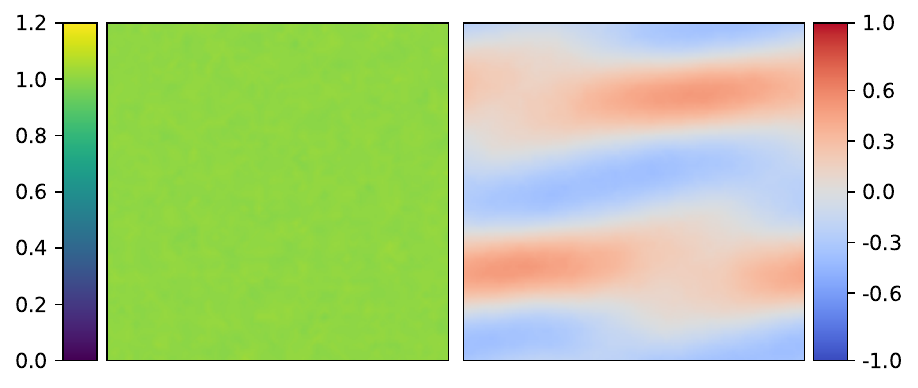}
(c) $r_s=35$
\end{minipage}
\begin{minipage}{0.48\textwidth}
\includegraphics[width=\linewidth]{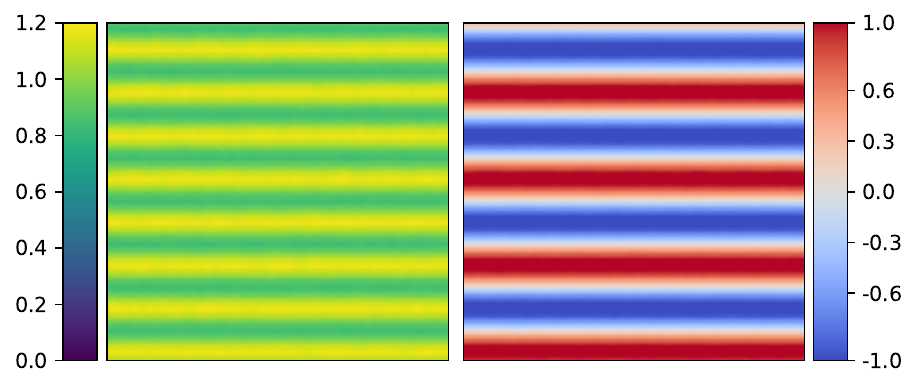}
(d) $r_s=45$
\end{minipage}
\caption{Evolution of the charge (left) and spin (right) densities. We choose the normalization so that the charge density averages to one in the simulation cell. }
\label{fig:evodens}
\end{figure}

\begin{figure}[!htbp]
\begin{minipage}{0.48\textwidth}
\includegraphics[width=\linewidth]{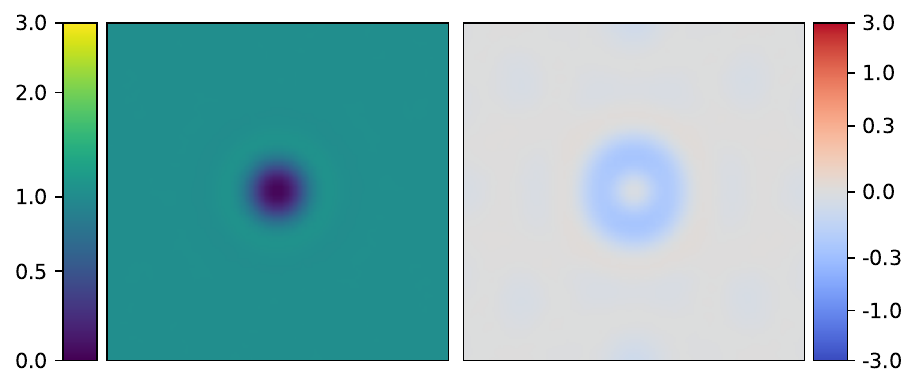}
(a) $r_s = 5$
\end{minipage}
\begin{minipage}{0.48\textwidth}
\includegraphics[width=\linewidth]{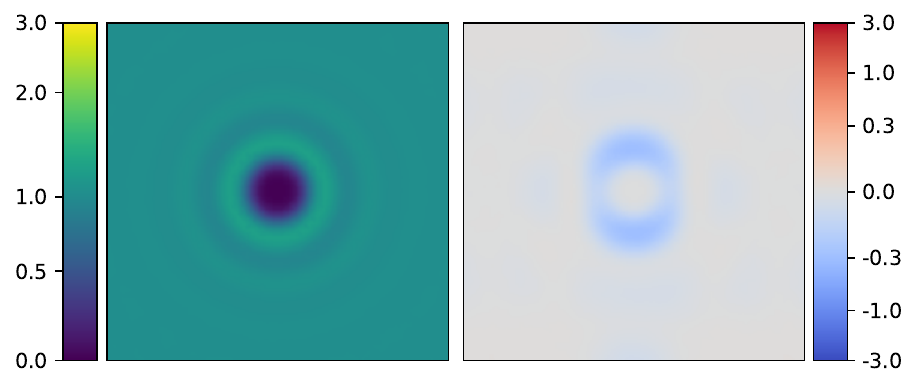}
(b) $r_s=20$
\end{minipage}
\begin{minipage}{0.48\textwidth}
\includegraphics[width=\linewidth]{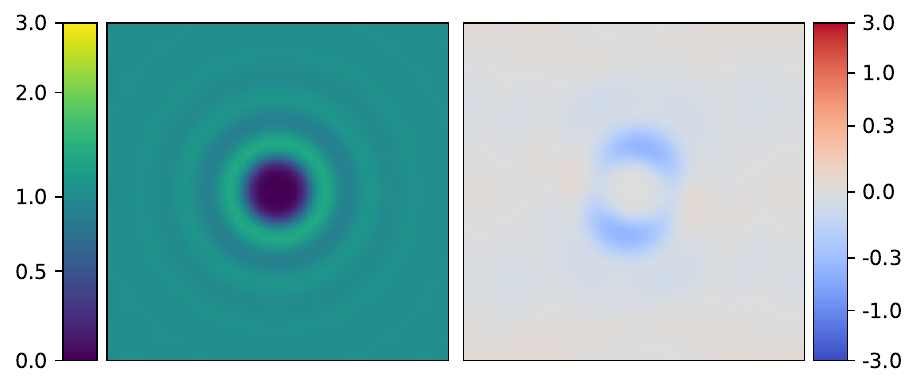}
(c) $r_s=35$
\end{minipage}
\begin{minipage}{0.48\textwidth}
\includegraphics[width=\linewidth]{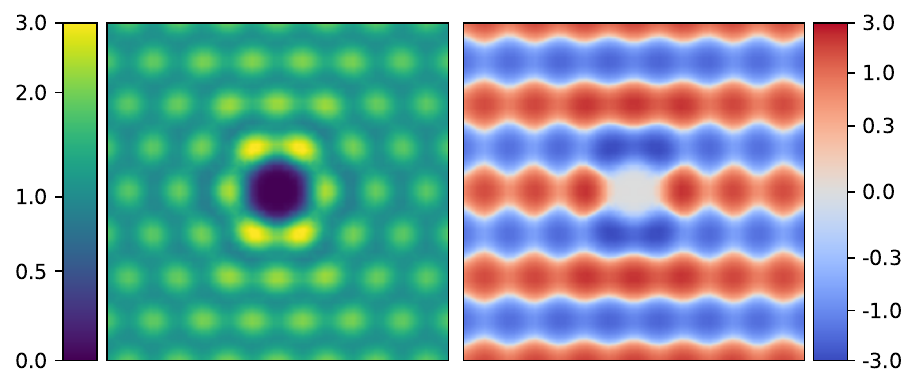}
(d) $r_s=45$
\end{minipage}
\caption{Evolution of pair correlation functions $g(r)$ (left) and spin-spin correlation function $g_s(r)$ (right). Note the colorbar is nonuniform in order to capture different scales between liquid and crystal phases.}
\label{fig:evogofr}
\end{figure}

\begin{figure}[!htbp]
\begin{minipage}{0.3\textwidth}
\includegraphics[width=\linewidth]{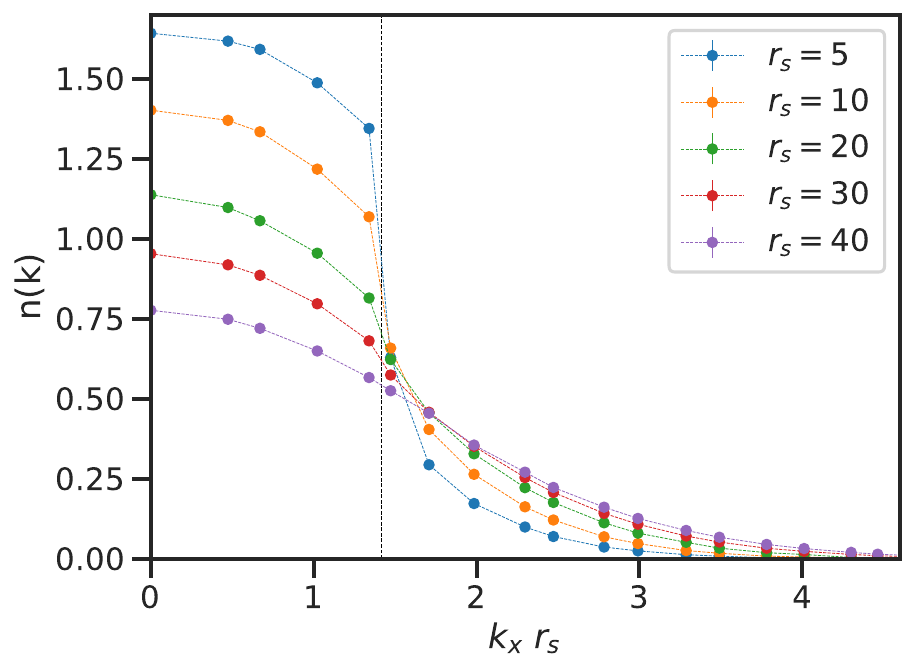}
(a) radial average
\end{minipage}
\begin{minipage}{0.3\textwidth}
\includegraphics[width=\linewidth]{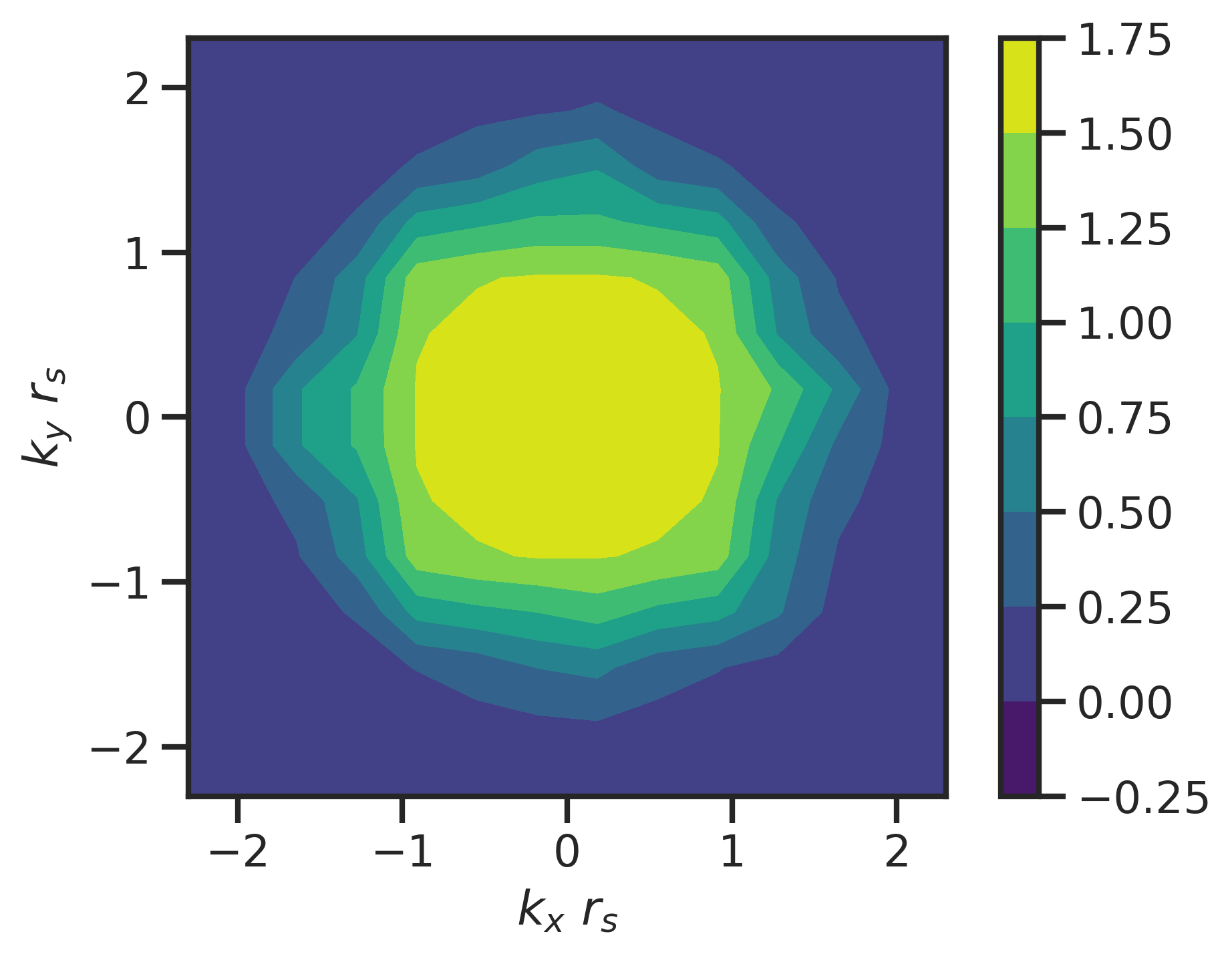}
(b) $r_s=5$
\end{minipage}
\begin{minipage}{0.3\textwidth}
\includegraphics[width=\linewidth]{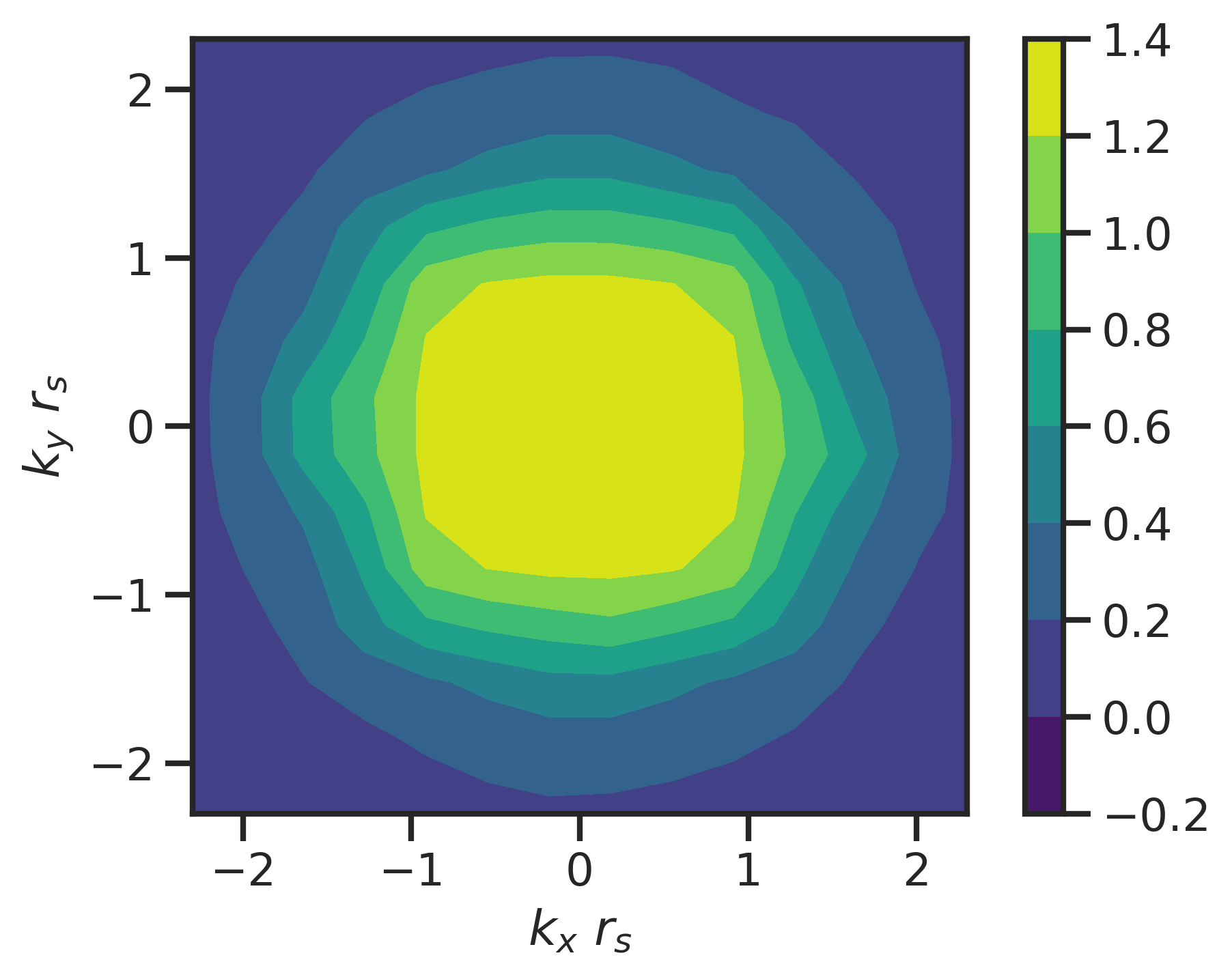}
(c) $r_s=10$
\end{minipage}
\begin{minipage}{0.3\textwidth}
\includegraphics[width=\linewidth]{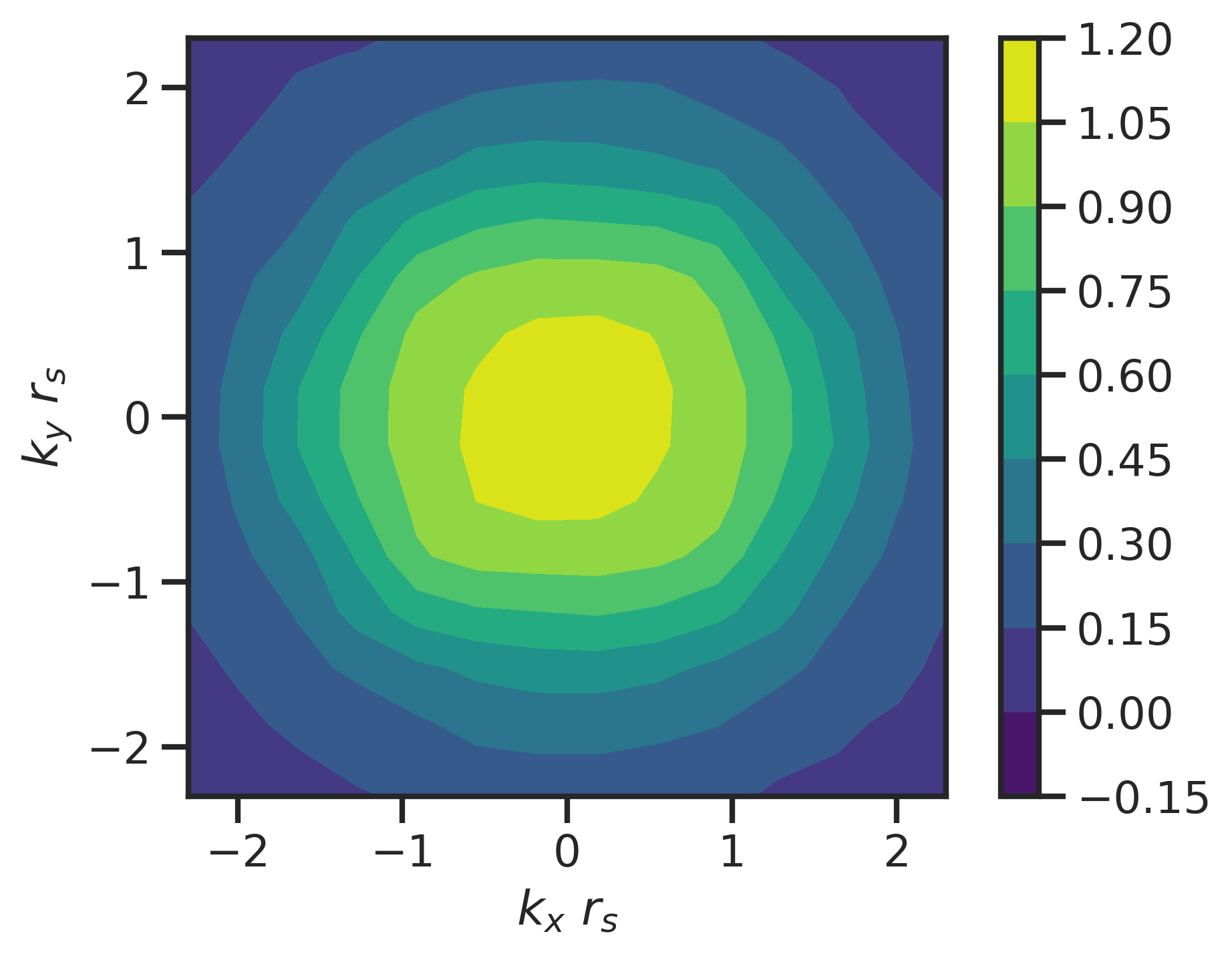}
(d) $r_s=20$
\end{minipage}
\begin{minipage}{0.3\textwidth}
\includegraphics[width=\linewidth]{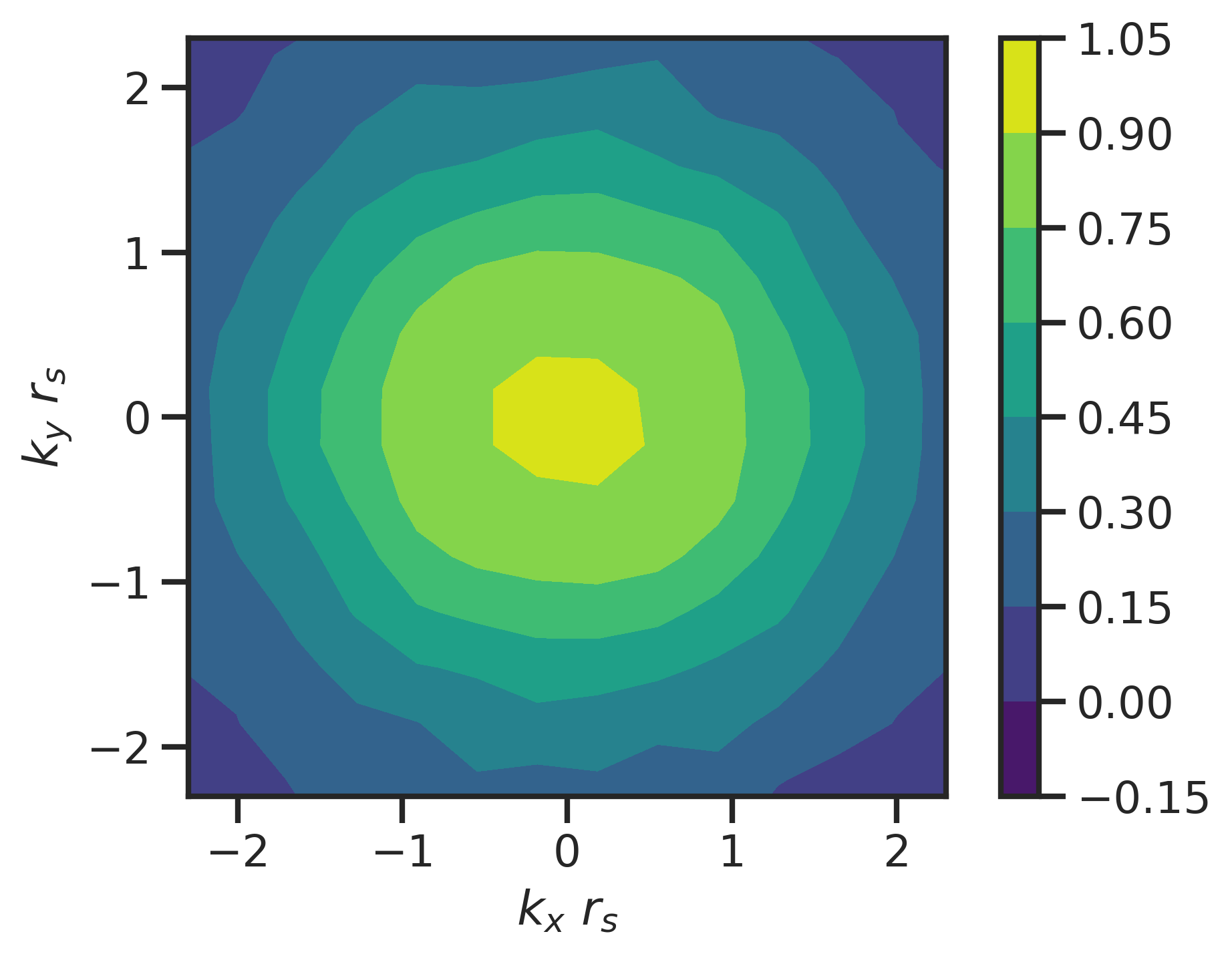}
(e) $r_s=30$
\end{minipage}
\begin{minipage}{0.3\textwidth}
\includegraphics[width=\linewidth]{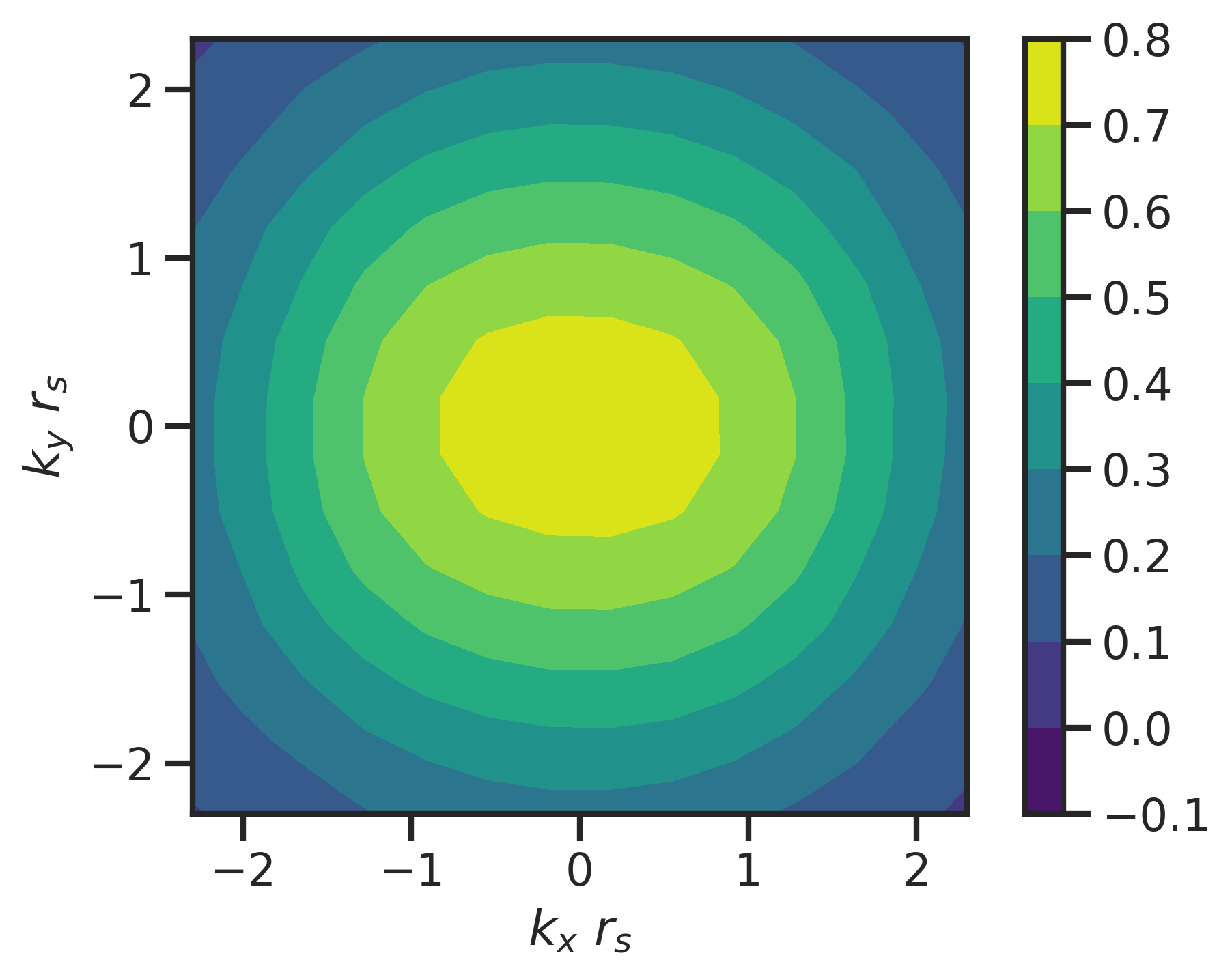}
(f) $r_s=40$
\end{minipage}
\caption{The momentum distribution $n(\bs{k})$ and its radial average. The Fermi surface is marked with a dashed black vertical line in (a).}
\label{fig:nofk}
\end{figure}

We show in this section, for various $r_s$, charge and spin densities in Fig.~\ref{fig:evodens}, charge and spin correlation functions in Fig.~\ref{fig:evogofr}, and momentum distributions
\begin{equation}
n(\bs{k}) = \dfrac{1}{N(2\pi)^2} \int d\bs{r}_0\int d\bs{r} e^{-i\bs{k}\cdot\bs{r}} \rho_2(\bs{r}_0, \bs{r}_0+\bs{r})
\end{equation}
in Fig.~\ref{fig:nofk}.
The charge density remains uniform in the liquid, whereas the spin density shows faint spin stripes in the \nematic~states.
The charge correlation remains isotropic in the liquid, where as the spin correlation shows short-range nematic AFM correlation in the \nematic~states.
The charge momentum distribution appears isotropic with a discontinuity at the Fermi surface $k=\sqrt{2}/r_s$ in the liquid.
The magnitude of the discontinuity decreases as $r_s$ increases and disappears in the WC phase.

\subsection{Anisotropy Observed in Square Simulation Cells}
\begin{figure}[!htbp]
\begin{minipage}{0.48\textwidth}
\includegraphics[width=\linewidth]{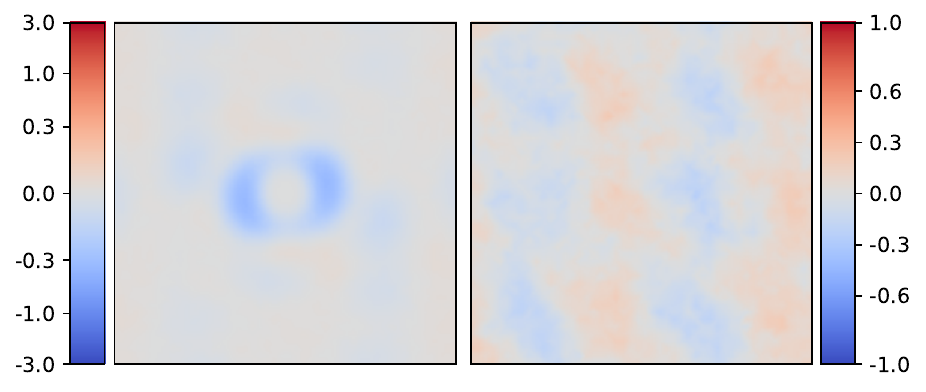}
(a) $r_s = 20$
\end{minipage}
\begin{minipage}{0.48\textwidth}
\includegraphics[width=\linewidth]{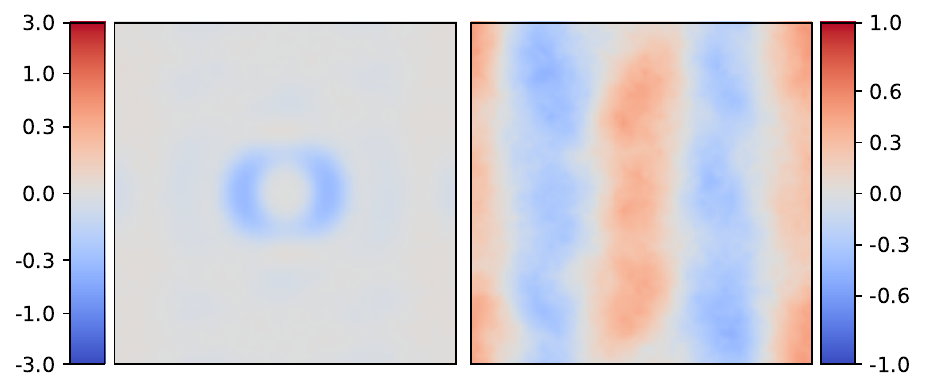}
(b) $r_s=25$
\end{minipage}
\begin{minipage}{0.48\textwidth}
\includegraphics[width=\linewidth]{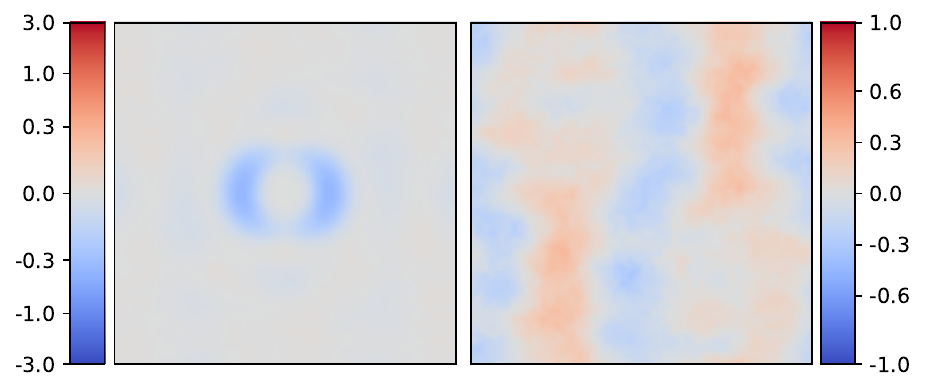}
(c) $r_s=30$
\end{minipage}
\begin{minipage}{0.48\textwidth}
\includegraphics[width=\linewidth]{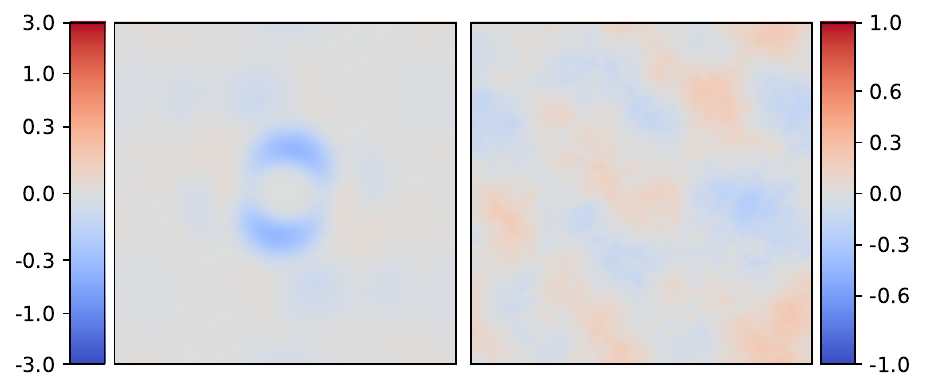}
(d) $r_s=35$
\end{minipage}
\caption{Spin-spin correlation $g_s(r)$ (left) and spin density (right) calculated for the lowest energy states at each $r_s$ in a square box with 58 electrons.}
\label{fig:square}
\end{figure}

In the main text, the rectangular simulation cell with $N=56$ electrons is constructed to be commensurate with a triangular lattice,
which could potentially induce an anisotropy in $g_s(r)$ and the one-body density patterns. %
To ensure the observed anisotropy is not an artifact of this cell choice, we also optimize the \ansatz~ansatz in a square cell with $N=58$ electrons.  
A Wigner crystal is frustrated in a square cell,
but similar anisotropy in $g_s(r)$ and the one-body density patterns as we approach $r_s$ of 35 can be seen in Fig. \ref{fig:square}.

\subsection{Hamiltonian Details}

The jellium Hamiltonian for $N$ electrons with mass $m$ in a uniform dielectric environment with constant $\epsilon$ is
\begin{equation} \label{eq:ham-ewald}
H = \dfrac{\hbar^2}{2m}\sum\limits_{i=1}^N \nabla_i^2 +
\dfrac{e^2}{4\pi\epsilon} V_E,
\end{equation}
where $V_E$ 
denotes the interaction energy:
\begin{equation}
V_E = %
\sum\limits_{i=1}^{N} \sum\limits_{j=i+1}^{N}
\sum\limits_{\bs{L}_s}
\dfrac{1}{\vert \bs{r}_i-\bs{r}_j-\bs{L}_s\vert} + b.g. \equiv
\sum\limits_{i=1}^N \sum\limits_{j=i+1}^N
v_E(\bs{r}_i-\bs{r}_j)
+\dfrac{1}{2}Nv_M,
\end{equation}
where $\bs{L}_s$ is the Bravais lattice generated by the simulation cell with periodic boundary condition.
A uniform compensating background is added to enforce charge neutrality.
The infinite sum over $\bs{L}_s$ defines the Ewald interaction
between a pair of point charges
$v_E(\bs{r})$ and the Madelung energy
\begin{equation}
v_M\equiv\lim_{\bs{r}\rightarrow\bs{0}}[v_E(\bs{r})-1/r],
\end{equation}
which is the Ewald potential of a single electron.
Direct evaluation of the infinite lattice sum over $\bs{L}_s$ is conditionally convergent, so a dual-space summation algorithm is typically used to evaluate the Ewald interaction
\begin{align}
v_E(\bs{r}) = \sum\limits_{\bs{L}_s} v^{sr}(\bs{r}-\bs{L}_s)
+\dfrac{1}{\Omega} \sum\limits_{\bs{G}_s\neq\bs{0}} v^{lr}_{\bs{G}_s}
-\dfrac{1}{\Omega}\lim\limits_{\bs{G}\rightarrow\bs{0}}v^{sr}_{\bs{G}},
\end{align}
where the Coulomb interaction is expressed as the sum of short-range $v^{sr}$ and long-range $v^{lr}$ pieces $1/r=v^{sr}(r)+v^{lr}(r)$ with $v^{sr}_{\bs{G}}$ and $v^{lr}_{\bs{G}}$ being their Fourier transforms. In two spatial dimensions, we have the following concrete expressions: $v^{lr}_G=2\pi/G~\text{erfc}(G/(2\alpha))$, $v^{sr}(r) = 1/r~\text{erfc}(\alpha r)$ and $\lim\limits_{\bs{G}\rightarrow\bs{0}}v^{sr}_{\bs{G}}=2\sqrt{\pi}/\alpha$, where $\alpha$ is a screening length in reciprocal space.
$\bs{G}_s$ is the reciprocal lattice of $\bs{L}_s$.
More detailed analysis of the Coulomb interaction with periodic boundary conditions can be found in Refs.~\cite{drummond_finite-size_2008,Holzmann2016-fsc}. 

The Hamiltonian can be simplified using Hartree atomic units.
The Bohr radius is determined by the ratio of the constants
$a_B = \dfrac{e^2}{4\pi\epsilon}/\dfrac{\hbar^2}{2m}$ with a vacuum value of $\left(\dfrac{e^2}{4\pi\hbar^2}\right)~\dfrac{m_e}{\epsilon_0}=0.529177210544$~\AA.
Using $a_B$ as length units, eq.~(\ref{eq:ham-ewald}) becomes
\begin{equation} \label{eq:ham-ewald-rs}
H = E_h\left( \dfrac{1}{2} \tilde{\nabla}_i^2 + \tilde{V}_E \right),
\end{equation}
where the Hartree energy $E_h=\dfrac{\hbar^2}{m a_B^2}$ with a vacuum value of $27.21138624598$ eV.

\subsection{Wave Function Architectural Details }

\def\cA{{\mathcal A}}
\def\cF{{\mathcal F}}
\def\cJ{{\mathcal J}}
\def\cN{{\mathcal N}}
\def\cU{{\mathcal U}}
\def\cV{{\mathcal V}}

\def\RR{\mathbb{R}}

\newcommand{\mr}[1]{\mathrm{#1}}

Our neural network backflow $\cN$ is based on the message passing network proposed in \cite{pescia2023message}. The main idea is to iteratively update two streams of one- and two-body information $\bs{h}^{(t)}_i \in \RR^{d_1}$ and $\bs{h}^{(t)}_{ij} \in \RR^{d_2}$ for each electron $i$ and pair $i$ and $j$, at each message passing layer $t = 1, \dots, T$. In each layer $t$, the two streams are combined with the ``visible'' features of the system, $\bs{v}_i$ and $\bs{v}_{ij}$, to update the message, $\bs{m}^{(t)}_{ij}$, between electron pair $i$ and $j$.
\begin{align}
    \bs{g}^{(t)}_i = &~ \Big[ \bs{v}_i,\  \bs{h}^{(t-1)}_i \Big] \, , \\
    \bs{g}^{(t)}_{ij} = &~ \Big[ \bs{v}_{ij},\  \bs{h}^{(t-1)}_{ij} \Big] \, , \\
    \bs{h}^{(t)}_i = &~ \cF_{1}^{(t)}
    \Big( \sum_{j \neq i} \bs{m}^{(t)}_{ij},\ \bs{g}^{(t)}_i \Big) + \bs{h}^{(t-1)}_i \, , \\
    \bs{h}^{(t)}_{ij} = &~ \cF_{2}^{(t)}
    \Big( \bs{m}^{(t)}_{ij},\ \bs{g}^{(t)}_{ij} \Big) + \bs{h}^{(t-1)}_{ij} \, ,
\end{align}
where $\cF_{h_1}^{(t)}$ and $\cF_{h_2}^{(t)}$ are multi-layer perceptrons (MLPs). We include skip connections~\cite{he2016deep} to facilitate optimization. Here, we set $\bs{h}_i^{(0)} = h_1$ and $\bs{h}_{ij}^{(0)} = h_2$ for all $i$ and $j$ to ensure permutation equivariance, where $h_1$ and $h_2$ are optimizable vectors that are independent of $i$ and $j$ and initialized to zero. 

In this work we choose the ``visible'' features as the following:
\begin{align}
    \bs{v}_i = &~ \varnothing \, , \\
    \bs{v}_{ij} = &~ \Big[ \cos(2\pi A^{-1} \bs{r}_{ij}),\ \sin(2\pi A^{-1} \bs{r}_{ij}),\ \norm{\sin(\pi A^{-1} \bs{r}_{ij})},\ s_{ij} \Big] \, ,
\end{align}
where $A$ is the cell tensor of the simulation box, $\bs{r}_{ij} = \bs{r}_i - \bs{r}_j$ is the relative displacement between electron pair $i$ and $j$, and $s_{ij} = 2 \delta_{s_i, s_j} - 1$ assigns $+1$ for parallel spins and $-1$ for antiparallel spins. We use the sine and cosine function as they are smooth and consistent with the periodic boundary condition~\cite{pescia_neural-network_2022}. Note that we take the one-body features $\bs{v}_i$ to be empty to explicitly conserve the translational and time-reversal (spin-flipping) symmetry in the construction of neural network backflow $\cN$. Additional information such as $s_i$ and (periodic wrapped) $\bs{r}_i$ can be included in $\bs{v}_i$ to break the symmetry if desired.

The message $\bs{m}^{(t)}_{ij}$ is computed from another MLP $\cF_{m}^{(t)}$ that takes the two-body features $\bs{g}_{ij}^{(t)}$ as input, and is reweighted by an attention matrix $\cA^{(t)}_{ij}$ which itself is a function of $\bs{g}^{(t)}_{ij}$ from all electron pairs:
\begin{align}
    \bs{m}^{(t)}_{ij} = &~ \cA_{ij}^{(t)} \Bigl( \Bqty{\bs{g}_{ij}^{(t)}} \Bigr) \odot \cF_{m}^{(t)}\Big( \bs{g}_{ij}^{(t)} \Big) \, , 
\end{align}
where $\odot$ denotes element-wise multiplication. The attention matrix $\cA_{ij}^{(t)}$ is computed with the particle attention mechanism proposed in~\cite{pescia2023message} with modified scaling and an extra linear layer to facilitate training:
\begin{align}
    \cA_{ij}^{(t)} = &~ \mr{Linear}^{(t)} \circ \mr{GELU} 
        \pqty{\frac{1}{\sqrt{N}} \sum_l^N \bs{q}_{il}^{(t)} \bs{k}_{lj}^{(t)}} \, , \\
    \bs{q}_{ij}^{(t)} = &~ W_q^{(t)} \cdot \bs{g}_{ij}^{(t)} \, , \\
    \bs{k}_{ij}^{(t)} = &~ W_k^{(t)} \cdot \bs{g}_{ij}^{(t)} \, ,
\end{align} 
where $W_q^{(t)},W_k^{(t)} \in \RR^{d_2 \times d_2}$ are optimizable weight matrices that act on the feature dimension, and the attention mixes the information on the dimension of electron pairs. The ``$\circ$'' symbol represents function composition (i.e., $(f \circ g)(x) = f(g(x))$).

All the MLPs denoted by $\cF$ ($\cF_{1}^{(t)}$, $\cF_{2}^{(t)}$ and $\cF_{m}^{(t)}$) contain a single hidden layer with GELU~\cite{hendrycks2016gaussian} activation function, i.e.,
\begin{align}
    \cF = \mr{Linear}_2 \circ \mr{GELU} \circ \mr{Linear}_1 \, .
\end{align}
In addition, although not written explicitly in the equations, we apply layer normalization~\cite{ba2016layer} to the inputs of all MLPs and attention layers ($\cF$ and $\cA_{ij}$) to stabilize optimization.

After the message passing iterations, the backflow displacement $\cN$ is computed by a linear transform from the final one-body stream $\bs{h}_i^{(T)}$:
\begin{align}
    \bs{x}_i = \bs{r}_i + \cN(\{\bs{r}_i\}) = \bs{r}_i + W_\mr{bf} \cdot \bs{h}_i^{(T)} \, .
\end{align}
where $W_\mr{bf} \in \RR^{d_1 \times 2}$ is another weight matrix that maps the one-body stream to the displacement of electron $i$ in two dimensional space.

Similarly, the extra term $\cU(\{\bs{r}_i\})$ in the Jastrow factor is a sum of one-body contributions, which are computed from the final one-body stream $\bs{h}_i^{(T)}$ and the backflow wrapped coordinates $\bs{x}_i$:
\begin{align}
    \cU(\{\bs{r}_i\}) = &~ \sum_i \cJ \pqty{\bs{h}_i^{(T)},\ \mr{GELU} \circ \mr{Linear_{pre}}(\bs{x}_i)} \, , \\
    \cJ = &~ \mr{Linear}_L \circ \mr{GELU} \circ \cdots \circ \mr{Linear}_1 \, ,
\end{align}
where $\mr{Linear_{pre}}:\ \RR^2 \rightarrow \RR^{d_1}$ is a linear layer that maps the wrapped coordinates $\bs{x}_i$ to a vector of the same dimension as one-body stream $\bs{h}_i$, so that they have comparable contribution in the MLP. $\cJ$ is a MLP with $L-1$ hidden layers that outputs a scalar. In practice, we use $L=4$ for $\cJ$ (i.e., 3 hidden layers). We also apply skip connections between all the hidden layers of $\cJ$ to facilitate optimization. 

In our calculations, we set the size of the hidden layers to 32 for all MLPs. We also set $d_1 = 32$ and $d_2 = 26$ so that the size of $\bs{g}_i$ and $\bs{g}_{ij}$ are equal to 32 as well. In addition, we choose the number of planewaves $N_k = 3N$ where $N$ is the number of electrons ($2N$ for 120 electron systems) rounding to the nearest closed shell value. This results in a total number of parameters slightly less than 50,000 for calculations with 56 electrons, which can be easily handled by our implementation of stochastic reconfiguration (SR). 

To summarize, our wavefunction ansatz largely follows the message passing network from~\cite{pescia2023message}, with the following major differences:
\begin{itemize}
    \item We use a linear combination of planewaves for each single-particle orbital $\phi_a$, instead of a single planewave or Gaussian function, to allow for automatic detection of liquid and crystal phases.
    \item We employ a dedicated Jastrow factor term $\cU$ that depends on both the one-body stream and the backflow coordinates, for better expressivity. We also include the standard pairwise term $u_2$ to help capture the cusp condition.
    \item We apply skip connection and layer normalization as well as some smaller modifications to facilitate optimization.
\end{itemize}

\subsection{Sampling and Optimization Details}

We employ the Metropolis-adjusted Langevin algorithm (MALA)~\cite{besag1994comments} to sample electron configurations  from the unnormalized probability given by the trial wavefunction $\abs{\Psi_T}^2$. Namely, at each step in the Markov chain, the new electron configuration $\tilde{R}$ is proposed by the following formula:
\begin{align}
    \tilde{R} = R + \tau \grad \log \abs{\Psi_T (R)}^2 + \sqrt{2 \tau} \xi,
\end{align}
where $R$ is the current electron configuration and $\xi$ is a random variable drawn from the standard Gaussian distribution in $2N$ dimension.
The proposed sample is then accepted or rejected according to the Metropolis-Hastings algorithm.
The step size $\tau$ is tuned adaptively during the optimization procedure to achieve a suitable acceptance rate. 

In practice, we run 1024 Markov chains in parallel, so that we have a batch size of 1024 in the estimation of energy and gradients. To reduce auto-correlation, we run 20 MALA steps between each optimization step, and only use the electron configurations generated from the last step in the optimization.
We increase or decrease the step size $\tau$ by 10\% every 10 optimization steps, to maintain the harmonic average of acceptance rate staying around $0.65$. We also clip the individual components of the gradient to have a absolute value smaller than one, to stabilize the acceptance rate in the sampling process. 
In both liquid and crystal phases, the initial electron configurations are chosen to be close to the positions of Wigner crystals with a small Gaussian perturbation. We thermalize the samples with 2000 MALA steps before the optimization starts. We find it is hard to change the spin pattern in the crystal phase in sampling, since it is a rare event for two localized electrons to exchange their position. 
Therefore we employ two strategy in initializing the spin configurations: aligning with the antiferromagnetic (AFM) pattern in the Wigner crystal or fully random. We find no difference for the two in liquid phase. In crystal phase, the AFM initialization generally achieves lower energy while the random initialization often gets stuck in local minima.

The optimization is conducted with a modified version of stochastic reconfiguration (SR) algorithm named SPRING~\cite{goldshlager2024kaczmarz}, which introduced an extra term, analogous to the momentum in stochastic gradient descent, to speed up convergence. Briefly speaking, the update of the parameters at step $t$ is given by
\begin{align}
    d\theta_t = &~ (S + \lambda I)^{-1}(g + \lambda\, \mu\, d\theta_{t-1}) \, , \nonumber \\
    \theta_{t+1} = &~ \theta_t - \eta\, d\theta_t
\end{align}
where $\theta_t$ stands for the parameters at step $t$. $S$ and $g$ are the quantum geometric (Fisher) matrix and gradients as in standard SR or natural gradient descent (NGD). $\eta$ is the learning rate, which is set to decay as a inverse function of $t$, $\eta = \eta_0 (1 + (t / T))^{-1}$ with $T$ being the ``delay'' of the decay. $\lambda$ and $\mu$ are hyperparameters controlling the strength of damping and momentum decay. Setting $\lambda = \mu = 0$ will reduce the algorithm to standard SR without diagonal shift. Following the technique in KFAC~\cite{martens2015optimizing}, we also scale $d\theta_t$ with a scalar so that its norm induced by the inverse Fisher matrix is smaller than a constant $C$. To efficiently solve the matrix inversion, we take advantage of the sparsity of Fisher matrix $S$ when the number of samples is smaller than the number of parameters, and apply the the Sherman-Morrison-Woodbury formula as described in Ref~\cite{ren2019efficient}.

In practice, we set $\lambda = 0.001$, $\mu = 0.9$, $T=1000$ and $C = \eta_0$. The starting learning rate is chosen dependent on density to accommodate different energy scales, as $\eta_0 = \{0.5, 2, 5\}$ for $r_s = \{5, 10, 15\}$, $\eta_0 = 10$ for $15 < r_s \leq 30$, and $\eta_0 = 15$ for $r_s > 30$ except that $\eta_0 = \{20, 25, 30\}$ for $r_s = \{35, 36, 50\}$.

\subsection{Visualizing the Optimized Wave Function}

To better understand the final trained wave function ansatz, we can attempt to visualize certain components independently. Through these components, there are signs of the liquid to crystal transition. However, as we are not sampling the wave function distribution, and are ignoring the determinant and Jastrow factor, these signals may be potentially spurious.

\begin{figure}[htbp]

\begin{minipage}{0.3\textwidth}
\includegraphics[width=\linewidth]{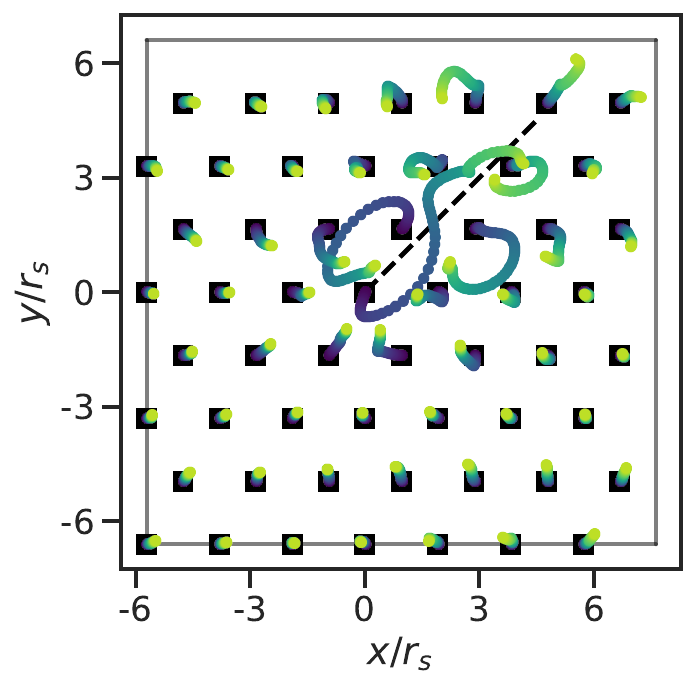}
(a) $r_s = 10$
\end{minipage}
\begin{minipage}{0.3\textwidth}
\includegraphics[width=\linewidth]{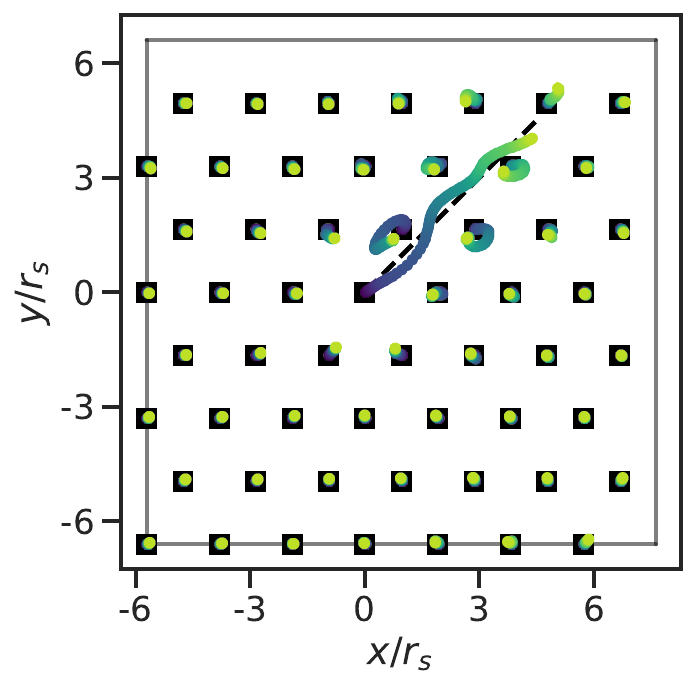}
(b) $r_s=34$
\end{minipage}
\begin{minipage}{0.3\textwidth}
\includegraphics[width=\linewidth]{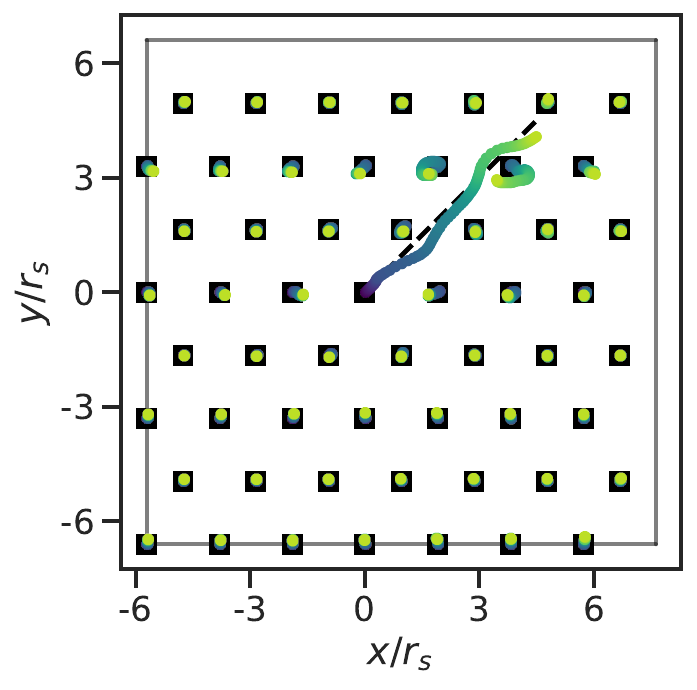}
(c) $r_s=45$
\end{minipage}

\caption{Final quasiparticle locations of the optimized backflow functions. The center electron is moved to a point along the black dashed line and quasi-particle positions are colored to correspond to electron initial displacement distance (lighter is further distance away). Initial WC positions are shown as black squares. }
\label{fig:app_backflow_traj}
\end{figure}

First we observe the effects of the backflow $\mathcal{N}$ by following the effects of displacing a single electron along a path, shown in Fig.~\ref{fig:app_backflow_traj}. 
For small $r_s$, such as shown in (a), there are strong corrections to the initial positions. For larger $r_s$, shown in (b) and (c) there is a much weaker effect and the electrons mostly remain at their initial positions, except when the displaced electron nears another electron. This matches the expected behavior seen in DMC, where backflow is essential to obtaining good liquid results at small to intermediate $r_s$. 

\begin{figure}[htbp]

\begin{minipage}{0.3\textwidth}
\includegraphics[width=\linewidth]{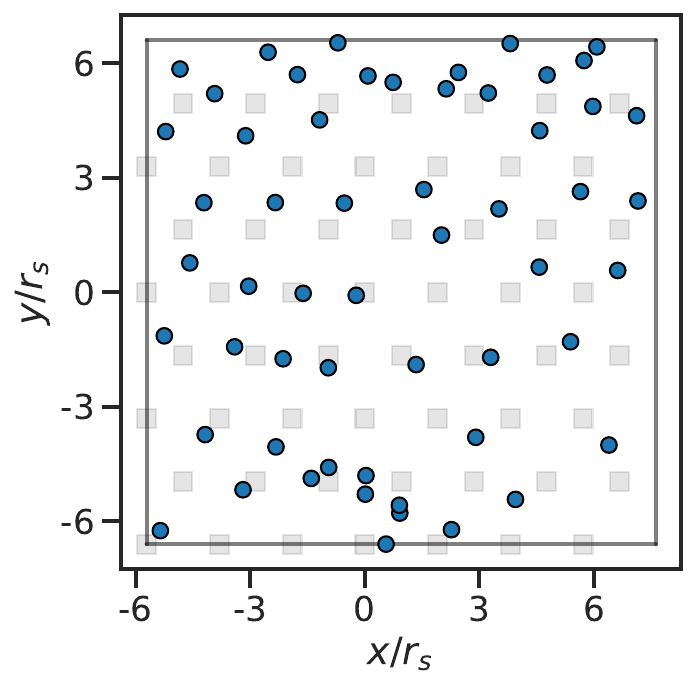}
(a) $r_s = 10$
\end{minipage}
\begin{minipage}{0.3\textwidth}
\includegraphics[width=\linewidth]{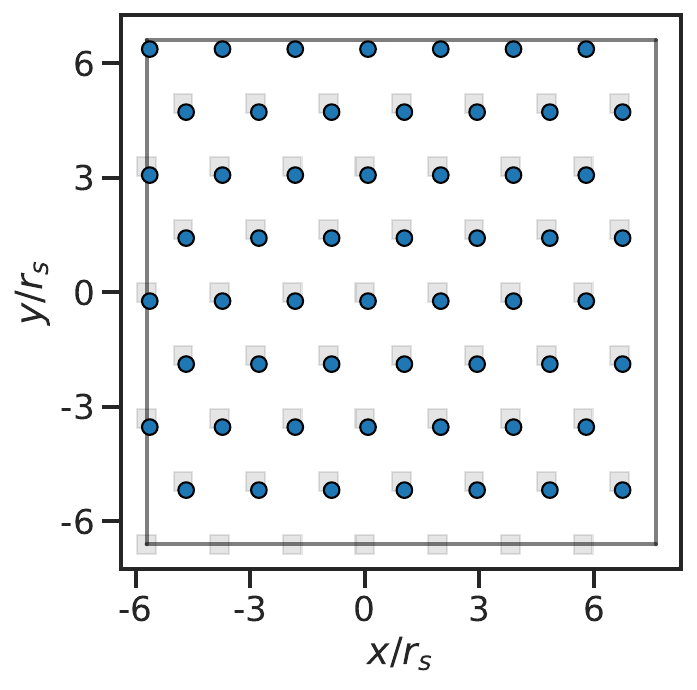}
(b) $r_s=34$
\end{minipage}
\begin{minipage}{0.3\textwidth}
\includegraphics[width=\linewidth]{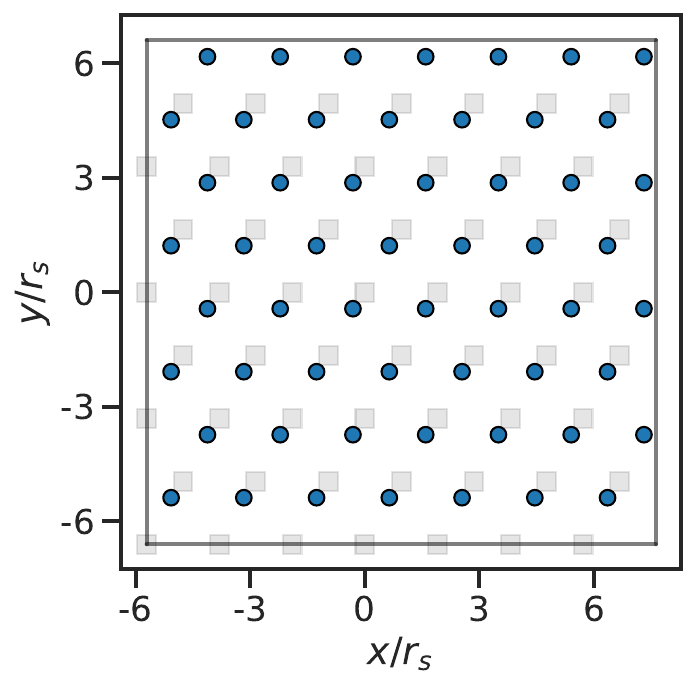}
(c) $r_s=45$
\end{minipage}

\caption{Final quasiparticle locations $\{r^{out}_j\}$ after iterive application of the optimized backflow function. The center electron is displaced a small amount diagonally and the resulting positions are recursively input 10 times with only the final positions plotted. Initial WC positions are shown as black squares.  }
\label{fig:app_backflow_traj_recursive}
\end{figure}

As another demonstration of backflow correlations, instead of displacing a single electron a varied distance, is to displace the electron once a small amount $\Delta r$ and recursively call the backflow on the output positions

\begin{equation}
    \{r^{out}_j\} = \mathcal{N} \circ \mathcal{N} \circ \dots \circ \mathcal{N}\left(\{x^c_j \cdot  (1+\Delta r \delta(r-r_0)\})\right),
\end{equation}
where $\{x^{c}_j\}$ have the electrons equi-spaced in a WC and $r_0$ denotes the electron at the origin.  After each application of $\mathcal{N}$, the positions are always projected back into the unit cell. 
Shown in Fig.~\ref{fig:app_backflow_traj_recursive} are the final quasiparticle positions after a displacement of $|\Delta r/r_s| \approx 0.06$ and iteratively applying backflow $\mathcal{N}$ to the resulting output 10 times. We see for small $r_s$ (Fig.~\ref{fig:app_backflow_traj_recursive}(a)), there is no apparent order to final positions. However for intermediate and large $r_s$ (Fig.~\ref{fig:app_backflow_traj_recursive}(b)-(c)), the output resembles the initial crystal positions having `corrected' the defect. Note that the shift from the initial positions can be removed by removing the center of mass movement. 

\begin{figure}[htbp]

\begin{minipage}{0.6\textwidth}
\includegraphics[width=\linewidth]{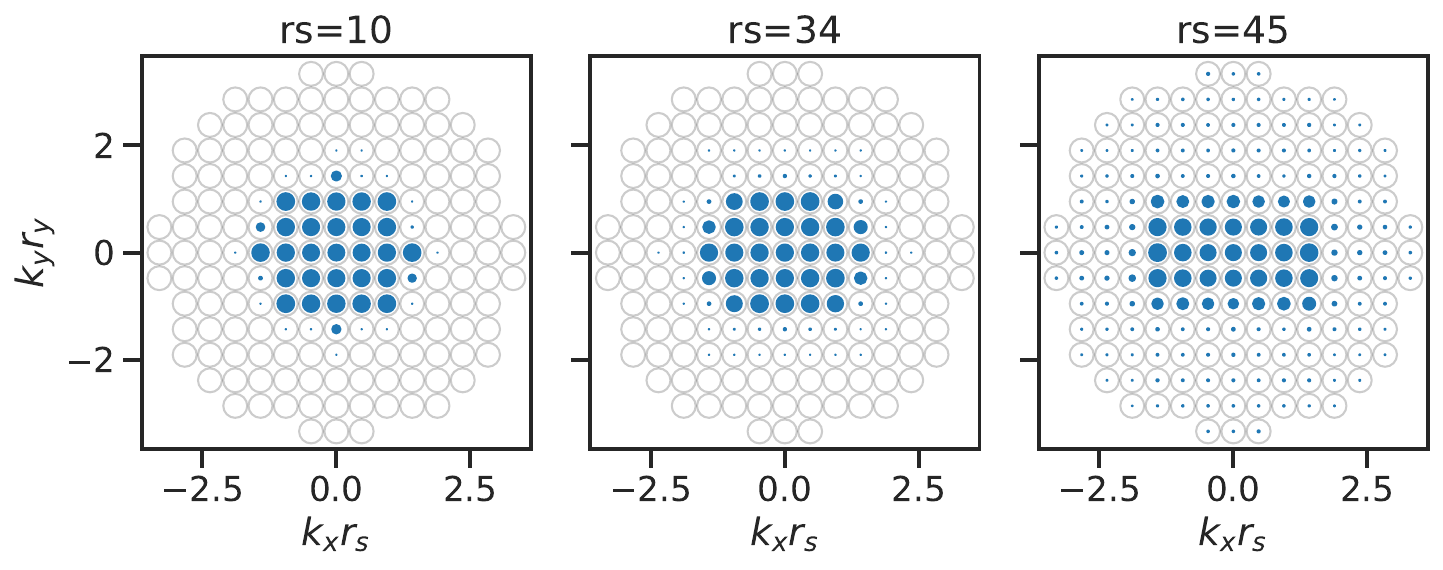}
(a) Spin Averaged Norm $|c_k|$
\end{minipage}
\begin{minipage}{0.3\textwidth}
\includegraphics[width=\linewidth]{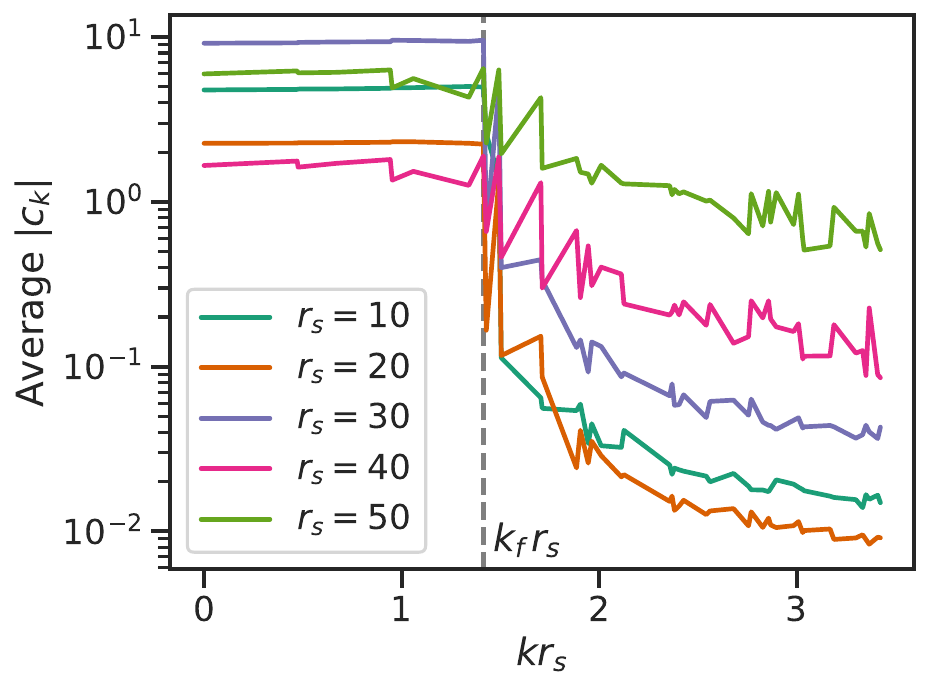}
(b) Spin and Radially Averaged Norm
\end{minipage}

\caption{Analysis of the optimized multiple planewave $\cmat^\sigma$ values.   }
\label{fig:app_C_mat}
\end{figure}

Then we can observe how the multi planewave coefficients $\cmat^\sigma$ are optimized across the different values of $r_s$, shown in Fig.~\ref{fig:app_C_mat}. First, the norm is taken over the electron index $a$ then averaged over spins ($\sigma=\{\uparrow\downarrow\}$) such that $|c_k| = 1/2 \sum_{\sigma} (\sum_{a} |\cmat^\sigma|^2)^{1/2} $. In Fig.~\ref{fig:app_C_mat}(a), each included $k$ value, a total of $N_k=3N$, is plotted as an open circle, while the magnitude of $|c_k|$ is shown as the relative to the maximum value. For low to intermediate $r_s$, only those $k$ points around $k_F$ are relatively strong although points further from $k_F$ pick up weight as $r_s$ increases. 
If we consider a threshold of $|c_k| \geq O(10^{-1})$ as significant, in the liquid phase, less than $40\%$ of the $k$ points reach that threshold but in the crystal phase nearly all of them do. This can be seen in Fig.~\ref{fig:app_C_mat}(b), where by radially averaging $|c_k|$, as $r_s$ increases the tail of the magnitudes at large $|k|$ slowly increases, until all $|c_k|$ are roughly on the same order.